\documentclass[trackchanges,preprint]{aastex63}
\usepackage{booktabs}
\usepackage{gensymb}
\usepackage[utf8]{inputenc}

\accepted{}
\turnoffeditone

\def\cataloglength{55} 
\def\knowncandidates{26} 

\begin{document}
\title{Monitoring the X-ray Variability of Bright X-ray Sources in M33}

\author[0009-0006-2411-5162]{Rebecca Kyer}
\affiliation{Department of Physics and Astronomy, Michigan State University, 567 Wilson Rd, East Lansing, MI 48824}
\affiliation{University of Washington Astronomy Department, Box 351580, Seattle, WA 98195, USA}

\author{Shelby Albrecht}
\affiliation{University of Washington Astronomy Department, Box 351580, Seattle, WA 98195, USA}

\author[0000-0002-7502-0597]{Benjamin F. Williams}
\affiliation{University of Washington Astronomy Department, Box 351580, Seattle, WA 98195, USA}

\author[0000-0002-0206-1208]{Kyros Hinton}
\affiliation{University of Washington Astronomy Department, Box 351580, Seattle, WA 98195, USA}

\author[0000-0002-4955-0471]{Breanna Binder}
\affiliation{Department of Physics \& Astronomy, California State Polytechnic University, 3801 W. Temple Ave, Pomona, CA 91768, USA}

\author[0000-0003-3252-352X]{Margaret Lazzarini}
\affiliation{Department of Physics and Astronomy, California State University Los Angeles, Los Angeles, CA 90032, USA}
\affiliation{Division of Physics, Mathematics and Astronomy, California Institute of Technology, 1200 E California Boulevard, Pasadena, CA 91125, USA}

\author[0000-0002-9202-8689]{Kristen Garofali}
\affiliation{NASA Goddard Space Flight Center, Code 662, Greenbelt, MD 20771, USA}

\author[0000-0003-2192-3296]{Bret Lehmer}
\affiliation{Department of Physics, University of Arkansas, 226 Physics Building, 825 West Dickson Street, Fayetteville, AR 72701, USA}

\author[0000-0002-3719-940X]{Michael Eracleous}
\affiliation{Department of Astronomy \& Astrophysics and Institute for Gravitation and the Cosmos, The Pennsylvania State University, 525 Davey Lab, University Park, PA 16802, USA}

\author[0000-0003-1415-5823]{Paul P. Plucinsky}
\affiliation{Center for Astrophysics $|$ Harvard \& Smithsonian, 60 Garden Street, Cambridge, MA 02138, USA}

\author[0000-0001-7539-1593]{Vallia Antoniou}
\affiliation{Department of Physics \& Astronomy, Texas Tech University, Lubbock, TX 79409, USA}
\affiliation{Center for Astrophysics $|$ Harvard \& Smithsonian, 60 Garden Street, Cambridge, MA 02138, USA}

\defcitealias{T11}{T11}
\defcitealias{M06}{M06}
\defcitealias{W15}{W15}
\defcitealias{G18}{G18}
\defcitealias{Lazzarini2023}{L23}

\begin{abstract}
We present a new five-epoch Chandra X-ray Observatory monitoring survey of the nearby spiral galaxy M33 which probes X-ray variability with time sampling between two weeks and four months. We characterize the X-ray variability of \cataloglength{} bright point sources outside of the nucleus, many of which are expected to be high-mass X-ray binaries (HMXBs). We detect eight new candidate transients not detected in previous X-ray catalogs of M33 and discuss their possible nature. The final catalog includes \knowncandidates{} known HMXB candidates identified in the literature. We extend the baseline of the X-ray light curves up to 21 years by including archival X-ray observations of these sources. We compare the detection and non-detection epochs of the sources to suites of simulated source duty cycles and infer that most of our detected sources have duty cycles $>30\%$. We find only four sources whose detection patterns are consistent with having duty cycles below $30\%$. This large fraction of sources with high duty cycles is unexpected for a population of HMXBs, thus more frequent X-ray monitoring will likely reveal many more low duty cycle HMXBs in M33.
\end{abstract}

\section{Introduction}

In an X-ray binary (XRB), a compact object accretes material from a stellar companion. Reaching this state requires a complex path of binary, and sometimes triple, star evolution \citep[e.g.,][]{Ford2000, Sana2012, Duchene2013,DeMarco2017}.  While our own Galaxy and its satellites contain many XRBs whose physical parameters have been measured in detail \citep{Liu2007, Antoniou2016, Haberl2016, Tetarenko2016}, statistical studies of the Milky Way population can be difficult due to our location in the dusty Galactic disk. By extending our studies of these sources to other galaxies where the distance to all XRBs is the same, we expand the sample to other environments and create homogeneous data sets that simplify statistical studies \citep[e.g.,][]{Prestwich2003, Tzanavaris2013,Haberl2016, Lehmer2021}. 

XRBs with donor stars $>8 M_{\sun}$ (high mass X-ray binaries; HMXBs) are of particular interest as they probe recent star formation within the galaxy \citep{Grimm2003,Ranalli2003,Gilfanov2004,Dray2006,Persic2007,Mapelli2010,Fragos2013a,Lehmer2017}. HMXBs have been found to contribute significantly to the overall X-ray emission in many galaxies \citep{Kim1992, Persic2002,Fabbiano2006,Mineo2012}, form heavy elements and distribute them to other parts of the host galaxy \citep{Ryu2016}, and may have been important contributors to cosmic reionization \citep{Kaaret2011,Mirabel2011,Fragos2013b,Mesinger2013,Brorby2014,Das2017,Heneka2020}.

X-ray variability is an important first-order indicator of the accretion mechanism driving HMXB emission. Multiple mass transfer mechanisms are possible depending on the configuration of the system, including stellar winds from the secondary star \citep[e.g.,][]{Hirai2021}, Roche-lobe overflow \citep[e.g.,][]{Chen2017}, passage of the compact primary through a Be-type companion's circumstellar disk \citep[e.g.,][]{Coe2015_circumstellardisk}, and wind-capture Roche-lobe overflow which may be responsible for powering the most luminous HMXBs \citep{Huarte-Espinosa2013, ElMellah2019}. Each case results in powerful X-ray emission originating from the compact object. Changes in an HMXB’s total X-ray emission are caused by changes in the accretion rate of material onto the compact object from the stellar companion. Several phenomenological X-ray source states for XRBs with a black hole primary describing differences in outburst behavior have been defined \citep{Homan2005, Tomsick2006, Kalemci2022}. These source states are described by luminosity as well as spectral and variability components, and transitions between them are linked to changes in the accretion mechanism \citep[e.g.,][]{RemillardMcClintock2006}. Although our survey is not sensitive enough to obtain spectra of these sources, changes in luminosity alone can indicate state changes. Variability studies over a wide range of timescales are critical to understanding these processes.

One important observational constraint on such variability is the duty cycle (DC), or ratio of time in outburst compared to the object's lifetime. The presence of highly variable sources with some percentage of time in a low X-ray luminosity state make it difficult to empirically determine the total number of HMXBs present in a galaxy \citep{Ducci2014,Haberl2016, Binder2017, Lazzarini2021, Mori2021}. The fraction of a galaxy’s HMXBs that always shows high emission (DC = 100\%) as a result of persistent accretion is uncertain. Collectively, measuring HMXB DCs is essential to our interpretation of the X-ray luminosity function (XLF) \citep{Belczynski2004,Paizis2014, Binder2017,Lehmer2019}. Within the last decade, the number of HMXB candidates and long-term X-ray observations has sufficiently grown to allow initial characterization of XRB DCs \citep{Binder2015,W15, G18, Lazzarini2021}. 

While individual HMXBs have been studied in detail across many wavelengths, there are few studies that analyze homogeneous populations of these objects and fewer that characterize their X-ray variability. The Small Magellanic Cloud (SMC) has provided one of the only well-studied populations to date \citep{Antoniou2019, Haberl2016, Antoniou2010}. Extensive studies of X-ray variability in the SMC sample have been performed \citep[e.g.,][]{Laycock2005,Galache2008, Haberl2016, Eger2008, Lazzarini2019, Haberl2022}. HMXBs have also been studied in the Large Magellanic Cloud (LMC) \citep{Lang1998, Coe2015_bexrb, Antoniou2016, vanJaarsveld2018, Maitra2023}, the dwarf starburst galaxy IC 10 \citep{Massey2007,Laycock2017a,Noori2017,Yang2019}, the spiral galaxy NGC 300 \citep{Kang2016, Binder2017}, and M31 \citep{Williams2004,Lazzarini2021}.

An additional nearby galaxy with a significant population of HMXBs is M33. At a distance of 817 kpc \citep{Freedman2001}, M33 is the second closest star-forming spiral galaxy. It lies at medium inclination \citep[$i=52 \degree$;][]{Kam2015} and has been well-observed across the electromagnetic spectrum (radio: \cite{Israel1974,Tabatabaei2022}; optical/UV: \cite{PHATTER2021}; $\gamma$-ray: \cite{Xi2020}). Previous X-ray surveys have localized point sources with high accuracy using the Chandra X-ray Observatory \citep[][\citetalias{T11} hereafter]{T11}, as well as with wide area coverage and several epochs observed by XMM-Newton \citep[][\citetalias{M06} and \citetalias{W15} hereafter]{M06, W15}. M33 has a high star formation rate, making its X-ray binary (XRB) population dominated by HMXBs \citepalias[e.g.][]{W15} which have secondary stars that are easily detected in HST imaging data \citep[][\citetalias{G18} hereafter]{G18,PHATTER2021}, and some with NuSTAR detections that allow a tentative classification of the compact object \citep{Yang2022}. \edit1{These characteristics make M33 an excellent target for studies of XRB variability.}

There have been five major surveys of M33 in the X-ray regime, three of which we utilize to study long-term variability of the sources in M33 between 2000-2015. M33 was first observed with the Chandra X-ray Observatory by \cite{Grimm2005}, and \cite{Grimm2007} used this survey to both characterize the population's X-ray variability and conduct spectral fitting. \citetalias{M06} conducted a deep XMM-Newton survey from 2000-2003. \citetalias{T11} conducted a deep Chandra survey from late 2005 to late 2006. \cite{W08} analyzed the \cite{Plucinsky2008} first look catalog of the \citetalias{T11} survey to identify seven X-ray transient candidates. \citetalias{W15} revisited M33 with XMM-Newton between 2010-2012. \cite{Yang2022} have also completed a NuSTAR survey of M33 in the hard X-ray band, classifying 28 sources as X-ray pulsars and BH binaries in various source states.

Pairing these surveys with optical data from the Hubble Space Telescope, \citetalias{G18} predicted $\sim$ 109 HMXBs in M33 assuming a star formation rate of 0.3 M$_{\sun}$ yr$^{-1}$ \citep{Williams2013a} and identified 55 HMXB candidates from archival HST fields that overlapped with the T11 survey. Mass and age estimates for 52 of these candidates have been obtained by \citetalias{G18} by analyzing their host stellar populations. The Panchromatic Hubble Andromeda Treasury: Triangulum Extended Region survey (PHATTER; \cite{PHATTER2021}) uniformly covers the inner disk of M33, including some regions not covered by the archival HST imaging used by \citetalias{G18}. \cite{Lazzarini2023} (\citetalias{Lazzarini2023} hereafter) identified 65 HMXB candidates by comparing the \citetalias{T11} X-ray positions to PHATTER optical counterparts consistent with being massive stars. \citetalias{Lazzarini2023} classified the companion spectral types and fit their spectral energy distributions to measure their physical properties. 

In addition to HMXBs, a shallow time-domain X-ray survey of M33 would be sensitive to strong outbursts from low-mass X-ray binaries (LMXBs). Accretion in a large fraction of these systems is attributed to the companion star filling its Roche-lobe (see \cite{Bahramian2023} for a recent overview). This accretion mechanism is highly efficient at converting energy into radiation and peak LMXB X-ray luminosities during outbursts are observed to range from $\sim10^{37}$ erg s$^{-1}$ to a few times $10^{38}$ erg s$^{-1}$ \citep{Yan2015} on timescales of days to months. These populations have been studied extensively in the bulge of M31 \citep[e.g.,][]{Williams2006,Peacock2010, Barnard2014}. While M33's XLF is dominated by HMXBs due to its younger stellar population \citep{Yang2022}, it is likely that some number of LMXBs are also present in the galaxy. Our survey would be sensitive to outbursts of these transient sources.

In this paper we present new Chandra observations of M33 taken over five epochs in 2021, nine years since \citetalias{W15} obtained the last X-ray observations of the galaxy. These new data allow us to characterize the long-term X-ray variability of \cataloglength{} bright sources, including \knowncandidates{} candidate HMXBs identified by \citetalias{T11}, \citetalias{G18}, \cite{Yang2022} and \citetalias{Lazzarini2023}. In Section 2, we describe the observations and data reduction. In Section 3, we present our catalog. In Section 4, we discuss constraints on the source DCs and how they compare to HMXB populations in other galaxies.

\section{Observations \& Data Reduction}
\begin{figure}[h]
\centering
\includegraphics[scale=0.8]{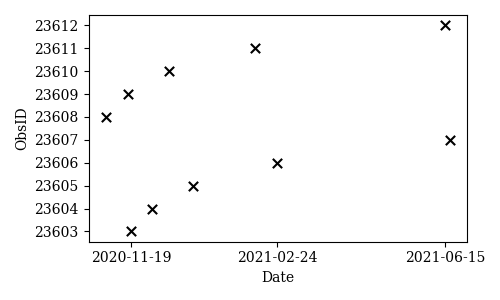}
\caption{Time sampling of ObsIDs in our survey. The lower points (ObsIDs 23603 to 23607) correspond to pointings centered on the Southern half of M33, and the upper points (ObsIDs 23608 to 23612) correspond to pointings centered on the Northern half of M33.}
\label{fig:timesampling}
\end{figure}

We obtained ten 5 ks observations of M33 with Chandra's ACIS-I detector. Five pointings were centered on the Northern half of M33 and five pointings were centered on the Southern half. Table \ref{table:obs_summary} gives the details of each observation. These ten exposures cover $\sim$66\% of the B-band D$_{25}$ isophote of M33 with observations taken between November 2020 and June 2021. The observations were designed to probe variability on multiple timescales. The five epochs in each pointing are separated by two weeks, four weeks, two months and four months consecutively. \edit1{The time sampling is shown in Figure \ref{fig:timesampling}.} We reprocessed all observations using CIAO v4.11 and CALDB v4.9.4 \citep{CIAO}. Below we describe our technique to measure our point source catalog and compare our catalog to previous surveys.

\begin{deluxetable*}{ccccccc}

\tablecaption{Summary of Chandra observations taken for this survey. The Pointing column indicates whether the ObsID was centered on the Northern (N) or Southern (S) halves of M33. The last column gives the effective exposure time after pipeline filtering.
\label{table:obs_summary}}

\tablehead{\colhead{Pointing} & \colhead{ObsID} & \colhead{Ob. Start} & \colhead{R.A. (J2000.0)} & \colhead{Dec. (J2000.0)} & \colhead{Roll Angle (\degr)} & \colhead{Exp. Time (ks)}}
\startdata
N & 23608 & 2020-11-02 & 23:32:14.463 & +30:45:10.75 & 222.84 & 4.99 \\
N & 23609 & 2020-11-17 & 23:32:16.673 & +30:45:03.02 & 253.17 & 4.99 \\
N & 23610 & 2020-12-14 & 23:32:20.110 & +30:44:58.83 & 277.05 & 4.99 \\
N & 23611 & 2021-02-09 & 23:32:25.525 & +30:45:01.15 & 298.87 & 4.81 \\
N & 23612 & 2021-06-15 & 23:32:36.830 & +30:45:20.96 & 91.19 & 4.99 \\
S & 23603 & 2020-11-19 & 23:27:30.881 & +30:35:54.15 & 256.37 & 4.99 \\
S & 23604 & 2020-12-02 & 23:27:33.631 & +30:35:52.27 & 269.82 & 4.99 \\
S & 23605 & 2020-12-30 & 23:27:39.216 & +30:35:49.92 & 297.20 & 4.99 \\
S & 23606 & 2021-02-24 & 23:27:40.504 & +30:35:52.40 & 304.78 & 4.99 \\
S & 23607 & 2021-06-18 & 23:27:47.535 & +30:36:14.47 & 99.48 & 4.99
\enddata

\end{deluxetable*}

\subsection{Source Detection and Point Source Extraction}\label{subsec:data_calibration}

Source detection was performed with 0.492 arcsec and 0.984 arcsec pixel sizes for completeness before pruning the sources based on the extracted properties. We used the CIAO tool {\tt fluximage} to create aspect histograms and point spread function files for each observation with both pixel sizes. We conducted source detection on each images with the CIAO tool {\tt wavdetect} on both pixel sizes and size scales ranging from 1 to 16. We produced separate source location lists for each of the two pixel sizes and merged them to create our candidate source position list. We examined the positions reported for each Observation ID (ObsID) image to ensure that all potential sources were included in the {\tt wavdetect} lists. We aligned all observations by matching the three brightest source detections in each observation with their literature counterparts in \citetalias{T11} using {\tt wcs\_match}. We then used the standard recipes for ACIS-Extract v2021feb4 \citep[AE hereafter;][]{Broos2010, AE2012, AE} to extract the counts from the locations identified by {\tt wavdetect} in four energy bands: soft (0.35-1.1 keV), medium (1.1-2.6 keV), hard (2.6-8.0 keV) and full (0.35-8.0 keV), chosen to match those in the \citetalias{T11} catalog. AE extracts source properties from each ObsID and merges them into a single multi-ObsID extraction with multiple improved position estimates. We visually inspected all position estimates produced for the data in SAOImageDS9 \citep[DS9 hereafter;][]{DS9} and chose the most accurate estimate to be the new source position. We ran AE a second time with the updated positions using the 0.492 arcsec pixel size for the final point source extractions.

\begin{figure*}
\centering
\includegraphics[width=\textwidth]{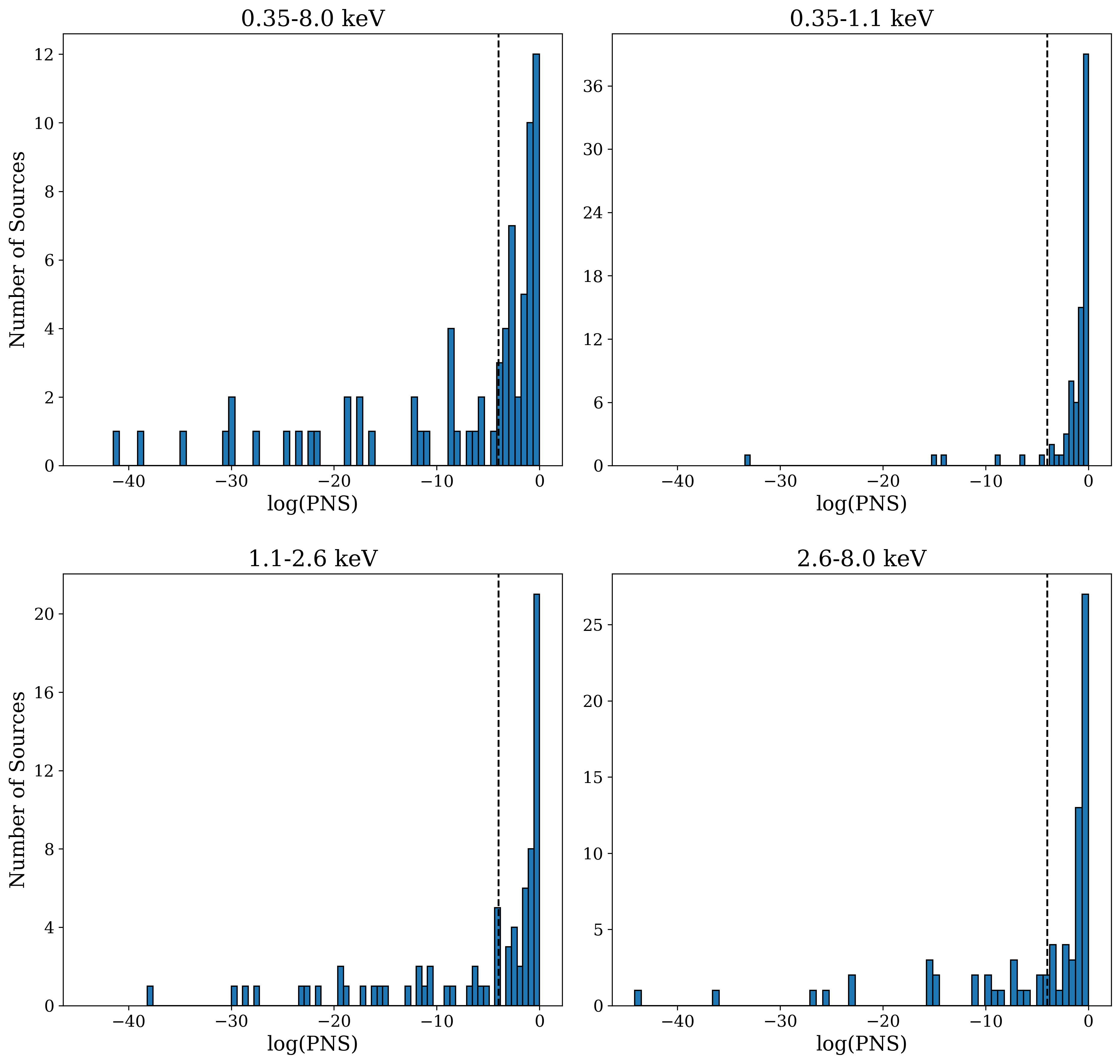}
\caption{Distribution of PNS values in the aligned multi-ObsID source extraction for each energy band. The vertical dashed lines indicate the maximum value of PNS = $10^{-4}$ used to select sources for the final catalog. Any source with PNS = 0 was omitted in the plot(s) for the bands in which that measurement occurred.}
\label{fig:PNS_distro}
\end{figure*}

We made final catalog selections based on the PROB$\_$NO$\_$SOURCE statistic (PNS) calculated by AE, which describes the probability of the extracted region not being a source. Figure \ref{fig:PNS_distro} shows that there were many sources in the initial list with high PNS in the multi-ObsID extraction, with a gap at about $10^{-10}$. We visually inspected the event lists of sources with PNS $>$ $10^{-5}$ in DS9 and compared the source lists based on cuts between PNS $> 10^{-8}$ and PNS $< 10^{-3}$. Our final catalog cut required PNS $< 10^{-4}$ (corresponding to $\gtrsim 2 \times 10^{-4}$ ct s$^{-1}$) in at least one energy band in either the multi- or single-ObsID extraction, to include sources with high variability. The final source list is insensitive to change down to PNS $> 5 \times 10^{-5}$ and includes all position matches to T11 (described in Section \ref{section:discussion}). For observations of some sources, AE reports zero or negative extracted counts, indicating that there was no detection during those particular observations.

\begin{figure}[h]
\centering
\includegraphics[width=0.5\textwidth]{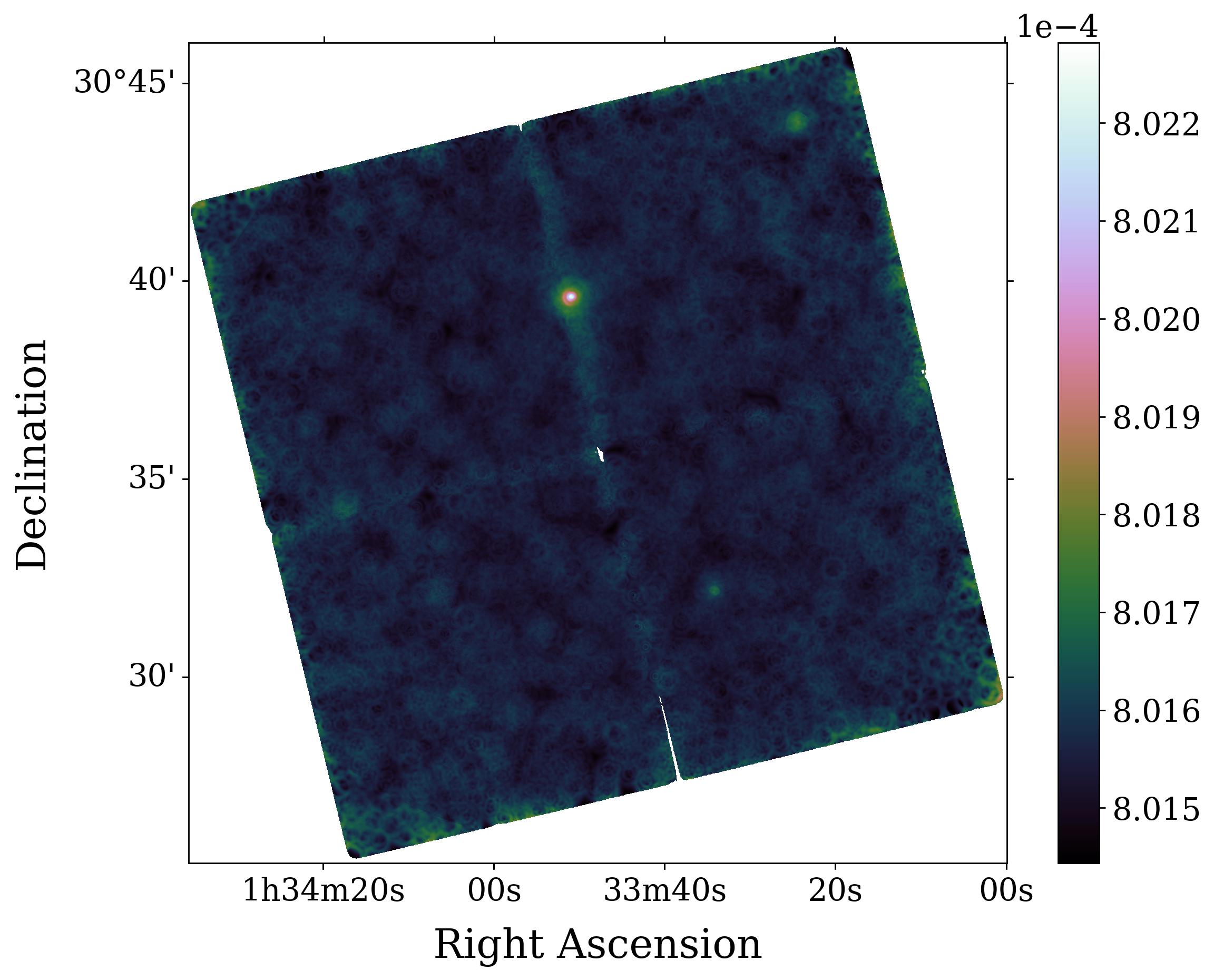}
\caption{Sensitivity map for ObsID 23603 in the 0.35-8.0 keV band. The colorbar is in units of ct s$^{-1}$. The feature in the chip gap is at the position of the bright nucleus source M33 X-8.}
\label{fig:sensmap}
\end{figure}

\begin{figure}[h]
\centering
\includegraphics[width=0.45\textwidth]{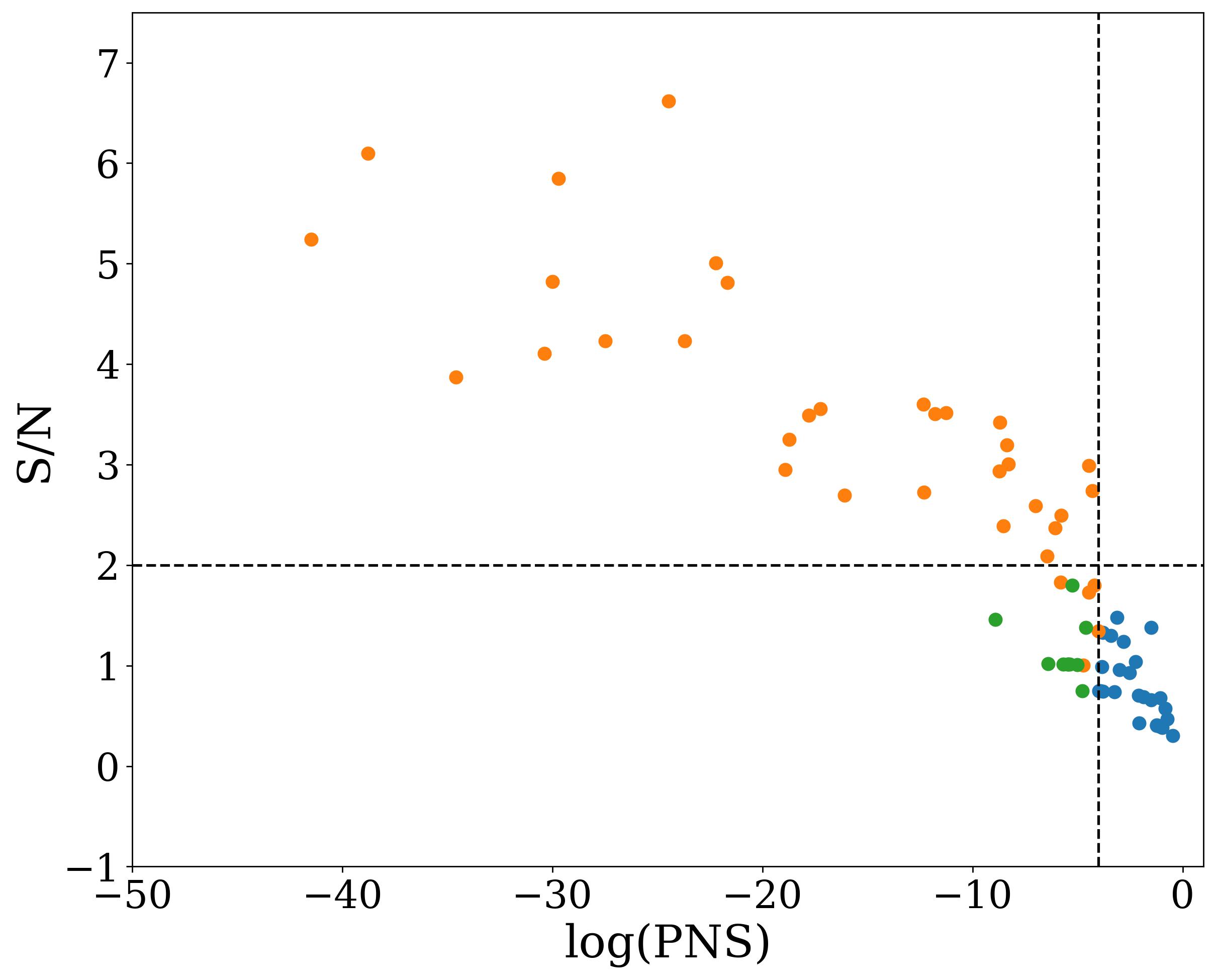}
\caption{Log of the signal to noise ratio (SRC\_SIGNIF reported by AE) plotted against the log of the PNS for all sources. The full band (0.35-8.0 keV) data is used if the source passes the PNS $<10^{-4}$ cut in the full band; otherwise the data is for the band where the PNS is lowest. Orange points are from the merged multi-ObsID extraction. Sources which only pass the PNS cut in an individual observation are represented by green points. Blue points show sources which do not pass the PNS cut in any observation or energy band, so the lowest PNS value between all merged and individual values is used. The horizontal dashed line indicates a signal to noise ratio of two. The vertical dashed line marks the PNS $=10^{-4}$ threshold used to select our final catalog as used in Figure \ref{fig:PNS_distro}.}
\label{fig:pnssigfin}
\end{figure}

To estimate the sensitivity of our survey we created sensitivity maps for each ObsID (see Figure \ref{fig:sensmap}) using the exposure corrected background images produced by {\tt wavdetect} with a cutoff signal to noise ratio (S/N) $>2$. The average of the median sensitivities in all ten observations is $8\times{10}^{-4}$ ct s$^{-1}$. Of the final catalog sources, 76.8\% have count rates larger than the average median sensitivity in the full 0.35-8.0 keV band, and \edit1{all source count rates exceed this sensitivity when considering their upper errors}. The S/N cutoff $>2$ aligns with sources that pass the multi-ObsID PNS cut as seen in Figure \ref{fig:pnssigfin}, and no sources with PNS $>$ $10^{-4}$ have S/N $>2$. 41 sources have a S/N $>2$ while 56 sources have PNS $< 10^{-4}$ due to using a flat S/N-based sensitivity which does not account for lower sensitivity at high off-axis angles. 

The nucleus of M33 contains the known ultraluminous X-ray source (ULX) M33 X-8 \citep{West2018,Krivonos2018}, which is known not be a transient source. We include this source in our tables for completeness but in this work we focus on the non-ULX HMXB population. We have excluded the nuclear source from all figures. Our observations may be of interest for future dedicated studies of M33 X-8. 

\begin{figure*}
\centering
\includegraphics[width=\textwidth]{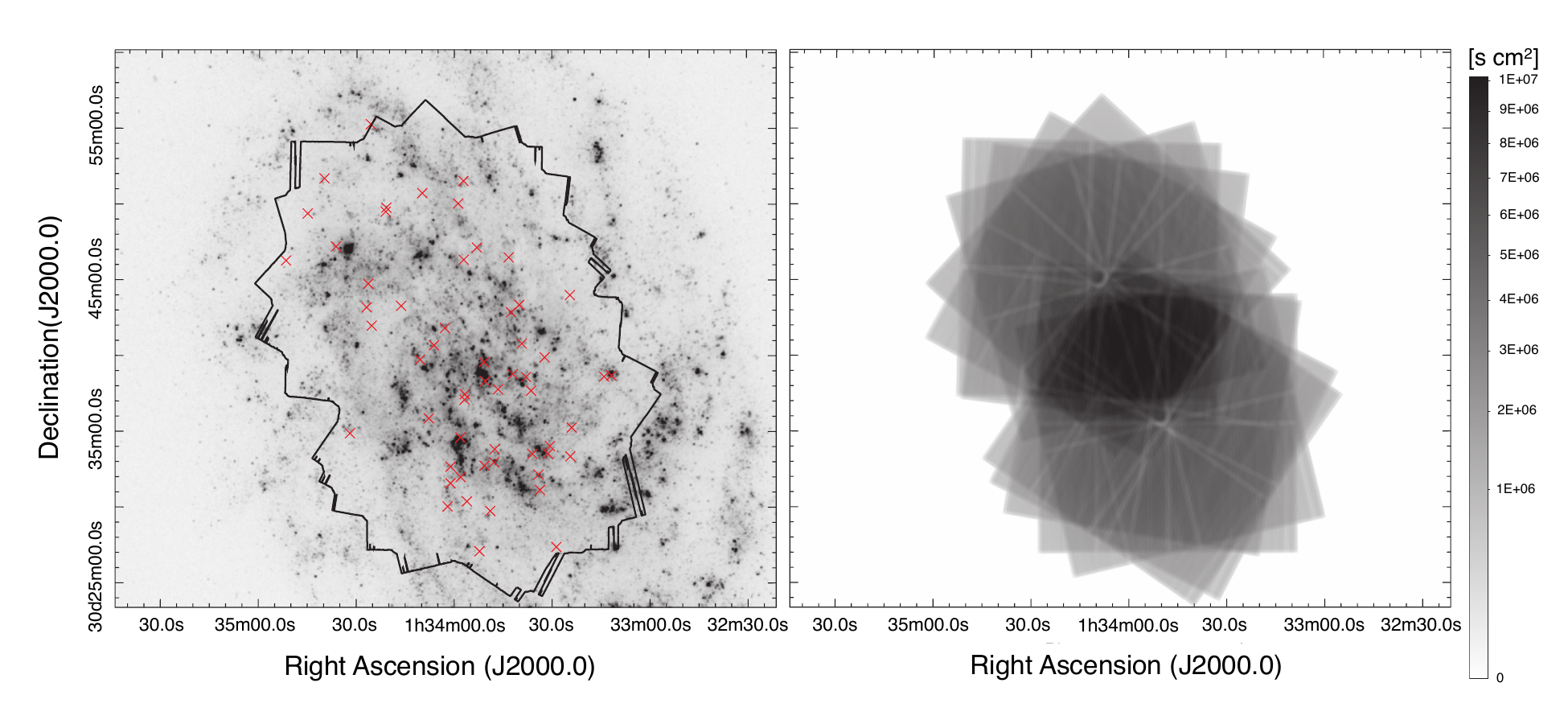}
\caption{Left panel: Outline of the total sky coverage of the ten pointings in this survey (solid black line) and positions of our catalog sources (red crosses) overlaid on a smoothed GALEX UV image of M33 \citep{GALEX_img}. Right panel: Stacked exposure maps on a square root scale in the 0.35-8.0 keV band.}
\label{fig:coverage}
\end{figure*} 

\subsection{Matching to Previous X-ray Surveys} \label{subsec:previousxraysurveys}

The rich observational history of M33 allows us to compare our observations to multiple archival X-ray surveys of the galaxy to study the variability of sources over a longer baseline, as well as to calibrate our astrometry. We cross matched our catalog to source positions in \citetalias{M06}, \citetalias{T11} and \citetalias{W15} using the catalog matching tool NWAY \citep{NWAY}. We used a 10 arcsec search radius for completeness of matches to faint off-axis sources in our catalog with high positional uncertainties. 38 sources in the catalog have matches to \citetalias{M06}, with a median offset of 1.3 arcsec and maximum offset of 6.6 arcsec. 46 sources in the catalog have matches to \citetalias{T11} with a median offset of 0.5 arcsec and maximum offset of 3.6 arcsec. 45 sources in our catalog have matches to \citetalias{W15} with a median offset of 1.0 arcsec and maximum offset of 5.3 arcsec. Sources with archival counterparts are indicated in Table \ref{table:final_catalog_summary} and the catalog source identifiers are given in the full table online. We discuss eight sources which do not have matches to any archival sources as new transients in Section \ref{section:newdetections}.

The soft-band sensitivity of the ACIS instrument on Chandra has been found to decrease over the observatory's lifetime \citep{Plucinsky2020}. AE reports flux estimates that take the instrument's response files into account (which are available in the full catalog table online), but for this shallow monitoring study we are primarily interested in whether or not a source is detected. To simplify comparison to archival observations, we found standard conversion factors between archival fluxes and modern Chandra count rates. We converted the archival fluxes of \citetalias{T11} to modern-day Chandra sensitivity using the web interface of the Portable, Interactive Multi-Mission Simulator (WebPIMMS; \cite{pimms}) by assuming a power law spectrum with photon index 1.7 and Galactic $N_H$ of $6.0\times10^{20}$ cm$^{-2}$, which are the same assumptions used by \citetalias{M06} and \citetalias{W15} for conversions between count rates and fluxes. We found a conversion factor of 0.608 between count rates observed in Chandra Cycle 7 when \citetalias{T11} observations were taken and Cycle 22 when our data was obtained. The fluxes measured by the two XMM-Newton surveys must also be converted to their equivalent Chandra count rate due to differences in the sensitivity of the two observatories. We again used WebPIMMS and found a conversion factor of 1 equivalent Chandra count s$^{-1}$ = $(6.1 \times 10^{10}) \times F_{XMM}$ for the two XMM-Newton surveys, assuming the same spectral model. 

We calculated the median residual count rate between the archival surveys and our data for sources with PNS $< 10^{-10}$ to estimate the uncertainty in our calibration. The median residual compared to \citetalias{T11} is -21\%. The median power law index for the 38 \citetalias{T11} sources with best fit power laws is $\Gamma = 1.65 \pm 0.17$, corresponding to a range of conversion factors between 0.589 to 0.646. About 20\% of the cross matched sources to \citetalias{T11} have soft power law spectral fits with $\Gamma > 1.83$, which accounts for some of the overall difference in calibration between our survey and \citetalias{T11}. The comparison between our measurements and T11 is similar to typical comparisons between X-ray data sets. For example, the median difference from \citetalias{W15} is 10.3\%, and for \citetalias{M06} it is 16\%.

\begin{deluxetable*}{ccccc}

\tablecaption{Sensitivities of archival surveys compared to in this paper. Limiting unabsorbed luminosity in each respective band converted to flux assuming a distance of 817 kpc to M33 and to modern Chandra count rate using the same assumptions described in Section \ref{subsec:previousxraysurveys}. The average single-ObsID sensitivity of our survey is given in the last row. We describe the calculation of this sensitivity in Section \ref{section:results}.} \label{table:upperlims}

\tablehead{\colhead{Survey} & \colhead{Limiting L (erg s$^{-1}$)} & \colhead{Energy band (keV)} & \colhead{Limiting F (erg cm$^{-2}$ s$^{-1}$)} & \colhead{Chandra count rate (ct s$^{-1}$)}}
\startdata
M06 & 1.0e+35 & 0.2---4.5 & 1.3e-15 & 6.84e-05 \\
T11 & 2.4e+34 & 0.35---8.0 & 3.0e-16 & 1.38e-05 \\
W15 & 4.0e+34 & 0.2---4.5 & 5.0e-16 & 2.73e-05 \\
This work & 6.6e+35 & 0.35---8.0 & 8.3e-15 & 4.00e-04
\enddata

\end{deluxetable*}

\section{Results}\label{section:results}

With these data, we provide a final catalog of 55 sources plus the nucleus, discuss eight new transients detected in this survey, and present X-ray light curves with our detections and archival observations. \edit1{The catalog positions and survey coverage area are shown in Figure \ref{fig:coverage}.}

We report a summary of the catalog in Table \ref{table:final_catalog_summary} with source positions and archival counterpart identifiers. Table \ref{table:catalog_extractions} gives source likelihood statistics (PNS, S/N) and count rates for the merged and individual ObsID extractions in all four energy bands. Errors on the extracted net counts are given at the 1$\sigma$ confidence level using the Gehrels approximation for Poisson statistics \citep{Gehrels1986} as implemented in AE. The full table online includes all extraction results reported by AE. Our final catalog includes \cataloglength{} point sources, eight of which do not appear in previous X-ray surveys and are discussed in Section \ref{section:newdetections}.

The limiting luminosity of our survey is determined by the lowest positive count rate in a single observation among all point sources in the catalog. The limiting count rate is
$4\times10^{-4}$ ct s$^{-1}$. Assuming a power law spectrum with photon index 1.7, and Galactic $N_H$ of $6.0\times10^{20}$ cm$^{-2}$, we used WebPIMMS to find a conversion of 1 ct s$^{-1} = 2.11 \times 10^{-11}$ erg cm$^{-2}$ s$^{-1}$, yielding a limiting unabsorbed flux in the full energy band of $8.32 \times 10^{-15}$ erg cm$^{-2}$ s$^{-1}$. Assuming a distance of 817 kpc to M33, the limiting luminosity of our survey is $6.6 \times 10^{35}$ erg s$^{-1}$. For sources with soft X-ray spectra the limiting luminosity is higher due to Chandra's decrease in soft-band sensitivity, as discussed above. 

\begin{deluxetable*}{cccccccc}
\tabletypesize{\footnotesize}
\tablecaption{Preview of the final source catalog. The full catalog online contains 56 sources. Theta describes the average off-axis angle of the source on the detector. T11 Source Numbers refer to the \citetalias{T11} Final Catalog designations, not to be confused with the \cite{Plucinsky2008} first look catalog source numbers. Positional uncertainties reported by AE as less than 0.5 arcsec were set to 0.5 arcsec to match the minimum astrometric uncertainty of the \citetalias{T11} survey which we used for alignment. \label{table:final_catalog_summary}}
\tablehead{\colhead{Source Name} & \colhead{R.A. (J2000.0)} & \colhead{Dec (J2000.0)} & \colhead{Positional Error} & \colhead{$\theta$} & \colhead{Num. Obs.} & \colhead{Archival Matches} & \colhead{T11 Source Number}\\ \colhead{ } & \colhead{($\mathrm{{}^{\circ}}$)} & \colhead{($\mathrm{{}^{\circ}}$)} & \colhead{($\mathrm{{}^{\prime\prime}}$)} & \colhead{($\mathrm{{}^{\prime}}$)} & \colhead{ } & \colhead{ } & \colhead{ }}
\startdata
013311.76+303843.1 & 23.299014 & 30.645307 & 0.93 & 9.7 & 6 & M06, T11, W15 & 102 \\
013314.01+303839.0 & 23.308376 & 30.644156 & 0.93 & 9.2 & 6 &  &  \\
013323.96+303517.1$^{a}$ & 23.349821 & 30.588081 & 0.50 & 5.8 & 5 & M06, T11, W15 & 154 \\
013324.37+303323.0 & 23.351552 & 30.556381 & 0.87 & 6.2 & 5 &  &  \\
013324.47+304401.6 & 23.351958 & 30.733790 & 0.50 & 9.7 & 6 & M06, T11, W15 & 158 \\
013328.70+302724.4 & 23.369565 & 30.456766 & 0.50 & 9.8 & 4 & M06, T11, W15 & 180 \\
013330.65+303404.1 & 23.377702 & 30.567805 & 0.50 & 4.7 & 5 & M06, T11, W15 & 195 \\
013331.24+303333.0 & 23.380183 & 30.559170 & 0.50 & 4.8 & 5 & M06, T11, W15 & 197 \\
013332.24+303955.4 & 23.384336 & 30.665396 & 0.72 & 6.6 & 7 & M06, T11, W15 & 210 \\
013333.70+303109.8 & 23.390413 & 30.519378 & 0.50 & 6.0 & 5 & M06, T11, W15 & 221 \\
013334.14+303211.1 & 23.392249 & 30.536427 & 0.50 & 5.2 & 5 & M06, T11, W15 & 225 \\
013336.04+303333.1 & 23.400163 & 30.559207 & 0.50 & 3.9 & 5 & T11, W15 & 237 \\
013336.42+303743.1 & 23.401751 & 30.628636 & 0.86 & 5.1 & 7 & M06, T11, W15 & 238 \\
013337.94+303837.4 & 23.408091 & 30.643736 & 0.69 & 5.6 & 8 & T11, W15 & 250 \\
013339.25+304049.7 & 23.413524 & 30.680480 & 0.50 & 6.5 & 10 & M06, T11, W15 & 262
\enddata

\tablenotetext{a}{Source detected in the 0.5'' pixel size wavdetect run only (see Section \ref{subsec:data_calibration}). Ten additional single-pixel size detections are indicated in the full table online.}

\end{deluxetable*}

\begin{longrotatetable}
\begin{deluxetable}{llccccccccc}
\tabletypesize{\scriptsize}

\tablecaption{Preview of the final catalog source likelhood and count rates in every energy band considered for each individual ObsID and the merged extraction. The first source is an example of a source primarily located in Southern pointing observations, and the second source is an example of one primarily located in Northern pointing observations. Blank entries appear where a source was not within the field of an ObsID. $1 \sigma$ upper limits are given when a source's position was on the detector for an ObsID but was not detected. The full version of this table can be found online, as well as an additional table with one source per row including all extracted source properties returned by AE including source significance and background levels.  \label{table:catalog_extractions}}

\tablehead{\colhead{Source Name} & \colhead{ObsID} & \colhead{Exposure} & \multicolumn{4}{c}{PNS} & \multicolumn{4}{c}{Count Rate (10$^{-3}$ ct s$^{-1}$)}
\\ \colhead{ } & \colhead{ } & \colhead{(s)} & \colhead{0.35-8.0 keV} & \colhead{0.35-1.1 keV} & \colhead{1.1-2.6 keV} & \colhead{2.6-8.0 keV} & \colhead{0.35-8.0 keV} & \colhead{0.35-1.1 keV} & \colhead{1.1-2.6 keV} & \colhead{2.6-8.0 keV}}
\startdata
013311.76+303843.1 & Merged & 29955 & 3.3e-25 & 1.2e-14 & 5.4e-22 & 0.47 & $2.3_{-0.3}^{+0.4}$ & $0.9_{-0.2}^{+0.2}$ & $1.4_{-0.2}^{+0.3}$ & $< 0.2$ \\
 & 23603 & 4991 & 0.21 & 1 & 0.16 & 0.42 & $< 0.8$ & $< 0.3$ & $0.2_{-0.2}^{+0.5}$ & $< 0.6$ \\
 & 23604 & 4991 & 2.6e-11 & 0.013 & 2.3e-12 & 0.14 & $2.6_{-0.7}^{+1}$ & $0.4_{-0.3}^{+0.5}$ & $1.9_{-0.6}^{+0.9}$ & $0.3_{-0.3}^{+0.5}$ \\
 & 23605 & 4994 & 2.7e-10 & 1.4e-05 & 4.2e-09 & 0.48 & $2.4_{-0.7}^{+0.9}$ & $0.8_{-0.4}^{+0.6}$ & $1.5_{-0.6}^{+0.8}$ & $< 0.5$ \\
 & 23606 & 4994 & 5.5e-09 & 0.00052 & 1.9e-10 & 1 & $2.0_{-0.7}^{+0.9}$ & $0.6_{-0.3}^{+0.6}$ & $1.6_{-0.6}^{+0.8}$ & $< 0.3$ \\
 & 23607 & 4991 & 1.2e-06 & 0.0038 & 2.2e-05 & 0.091 & $1.5_{-0.6}^{+0.8}$ & $0.4_{-0.3}^{+0.5}$ & $0.8_{-0.4}^{+0.6}$ & $0.3_{-0.3}^{+0.5}$ \\
 & 23608 &  &  &  &  &  &  &  &  &  \\
 & 23609 &  &  &  &  &  &  &  &  &  \\
 & 23610 &  &  &  &  &  &  &  &  &  \\
 & 23611 &  &  &  &  &  &  &  &  &  \\
 & 23612 & 4994 & 4e-07 & 3.2e-09 & 0.00011 & 0.9 & $5.2_{-1}^{+2}$ & $3.6_{-0.9}^{+1}$ & $2.2_{-0.8}^{+1}$ & $< 0.09$ \\
\hline
013436.36+304713.9 & Merged & 34772 & 0.0002 & 0.71 & 0.14 & 3.4e-05 & $0.8_{-0.3}^{+0.3}$ & $< 0.08$ & $0.1_{-0.1}^{+0.2}$ & $0.7_{-0.2}^{+0.2}$ \\
 & 23603 &  &  &  &  &  &  &  &  &  \\
 & 23604 &  &  &  &  &  &  &  &  &  \\
 & 23605 & 4994 & 0.35 & 0.55 & 0.43 & 0.39 & $< 2$ & $< 0.7$ & $< 0.9$ & $< 1$ \\
 & 23606 & 4994 & 0.09 & 0.84 & 0.68 & 0.0087 & $1.5_{-1}^{+1}$ & $< 0.3$ & $< 0.6$ & $2.0_{-1}^{+1}$ \\
 & 23607 &  &  &  &  &  &  &  &  &  \\
 & 23608 & 4991 & 3.1e-05 & 0.059 & 0.09 & 0.0011 & $0.9_{-0.4}^{+0.7}$ & $0.2_{-0.2}^{+0.5}$ & $0.2_{-0.2}^{+0.5}$ & $0.6_{-0.3}^{+0.6}$ \\
 & 23609 & 4994 & 2.4e-05 & 1 & 1 & 1.8e-06 & $0.9_{-0.4}^{+0.7}$ & $< 0.4$ & $< 0.4$ & $1.0_{-0.4}^{+0.7}$ \\
 & 23610 & 4994 & 0.037 & 1 & 1 & 0.014 & $0.3_{-0.3}^{+0.5}$ & $< 0.4$ & $< 0.4$ & $0.4_{-0.3}^{+0.5}$ \\
 & 23611 & 4811 & 2.7e-05 & 1 & 0.0023 & 0.0013 & $1.0_{-0.4}^{+0.7}$ & $< 0.4$ & $0.4_{-0.3}^{+0.6}$ & $0.6_{-0.3}^{+0.6}$ \\
 & 23612 & 4994 & 0.0033 & 1 & 0.065 & 0.019 & $0.5_{-0.3}^{+0.6}$ & $< 0.4$ & $0.2_{-0.2}^{+0.5}$ & $0.4_{-0.3}^{+0.5}$
\enddata
\end{deluxetable}
\end{longrotatetable}

\begin{figure*}
\centering
\includegraphics[scale=.7]{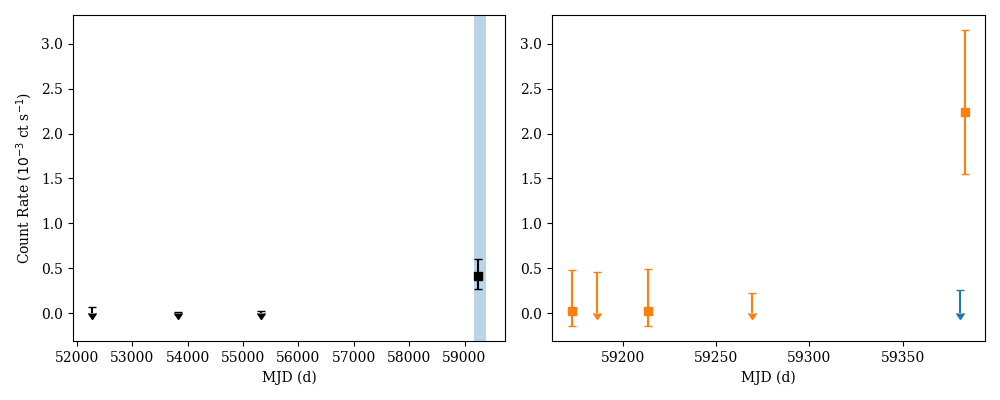}
\caption{X-ray light curve of Source 013314.01+303839.0, a new candidate transient detected by our survey with the highest net counts. The left panel shows the long-term light curve with archival upper limits from the \citetalias{M06}, \citetalias{T11} and \citetalias{W15} surveys along with our merged-ObsID count rate. The blue shaded region indicates the duration of time covered by our survey compared to the full baseline. The right panel shows the short-term light curve with count rates from individual ObsIDs within our survey only. Observations from pointings centered on the Northern half of M33 are marked in blue, and observations from pointings centered on the Southern half of M33 are marked in orange. Upper limits are shown with downward arrows where the source position was on the detector but no counts above the background were extracted.}
\label{fig:013313.95+303838.2_lc}
\end{figure*}

\begin{figure*}
\centering
\includegraphics[scale=.7]{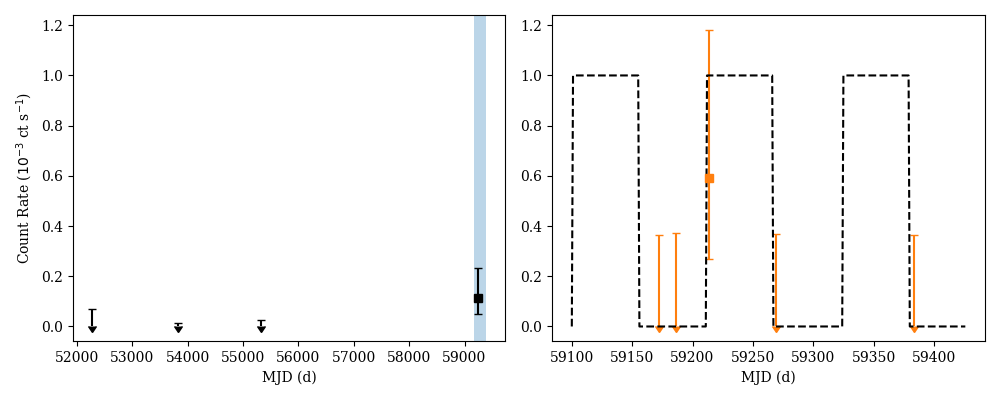}
\caption{X-ray light curve of Source 013347.76+303300.0, a new candidate transient detected by our survey with a duty cycle constrained by our simulations between 10-60\% on an outburst recurrence timescale between 10-200 days. The dashed black line on the right panel shows an example iteration of the simulated DC and timescale combination with the peak match to the observed light curve (DC = 50\% on 110 day timescale). The off times are represented at zero ct s$^{-1}$ and the on times are represented at $1 \times 10^{-3}$ ct s$^{-1}$.}
\label{fig:013347.77+303259.8_lc}
\end{figure*}
\clearpage
\subsection{X-ray Light Curves}

We show two examples of X-ray light curves for our catalog sources in Figures \ref{fig:013313.95+303838.2_lc} and \ref{fig:013347.77+303259.8_lc}. Light curves for all of the sources are included as supplemental data to the article. Long-baseline light curves incorporating the archival detections and upper limits from \citetalias{M06}, \citetalias{T11}, and \citetalias{W15} and our merged count rate are shown in the left panels. The duration of our monitoring campaign is indicated by the shaded region. Our survey's sky coverage lies within the exposure area of each of the archival catalogs, so in cases where our sources do not have archival counterparts we take the limiting sensitivity of each survey as the upper limit on the source's flux at that time. The archival upper limits used are given in Table \ref{table:upperlims}.

Short-term light curves using our own observations alone are given in the right panels. Observations from Northern and Southern pointings are marked by different colors. Nondetections when the source position was covered by the observation are represented in the light curve figures as downward arrows to zero ct s$^{-1}$. We find the upper limits for these points by adding the upper error to the extracted count rate. The extracted count rates are unchanged in the reported photometry online.

\subsection{New Candidate Transients \label{section:newdetections}}

\edit1{Galactic transient XRBs in quiescent states have X-ray luminosities less than $10^{33}$ erg s$^{-1}$ and reach values from $10^{36}$ to $10^{39}$ erg s$^{-1}$ in outburst \citep{Degenaar12}, most of which peak around $5 \times 10^{36}$ erg s$^{-1}$ \citep{Binder2017}. Given our detection limit of ~$6.6 \times 10^{35}$ erg s$^{-1}$, we expect to detect XRB sources in outburst and high-luminosity active accretion states only.}

Within our catalog, we find eight sources to be new X-ray detections with no counterparts in archival X-ray catalogs of M33. Each new \edit1{candidate} transient has a strong detection in at least one ObsID as summarized in Table \ref{table:new_transients}, and a few have positive extracted NET\_CNTS in a second ObsID that are close to our detection limit. We calculated minimum flux ratios between the detected count rate with the highest $1\sigma$ lower limit and the \citetalias{T11} survey sensitivity, which was the deepest archival survey and covered all source positions in our catalog. Previous studies define candidate X-ray transients as sources with a minimum flux ratio between five and ten \citep[e.g.][]{W08,Laycock2017a}. Seven of our eight previously undetected sources have minimum flux ratios greater than ten, and one source is just below this threshold. The minimum flux ratios for these new candidate X-ray transients are given in Table \ref{table:new_transients}. We stress that these are candidate transients because the large uncertainties on these source fluxes near our relatively shallow detection limit can lead to an overestimate of the number of transient sources in the catalog. \cite{Brassington2012} address this issue in their study of LMXB transients by utilizing a Bayesian approach to determine the minimum flux ratios with uncertainty estimates on the number of transient sources. If we take a more conservative approach to calculate the minimum flux ratio and divide the lower limit on the maximum detected count rate by the upper limit on the nondetection survey sensitivity, then three of the eight sources have minimum flux ratios greater than five (013314.01+303839.0, 013324.37+303323.0, and 013406.22+304042.9). Nevertheless, the low background of the ACIS instrument and uncrowded field still make these detections possible transient sources, particularly the three sources which are detected primarily in the soft 0.35---8.0 keV band despite its degraded sensitivity. The variability characteristics and DC constraints for these sources are discussed in Sections \ref{subsec:variability} and \ref{subsec:simdutycycles}.

\begin{figure}
\centering
\includegraphics[width=0.47\textwidth]{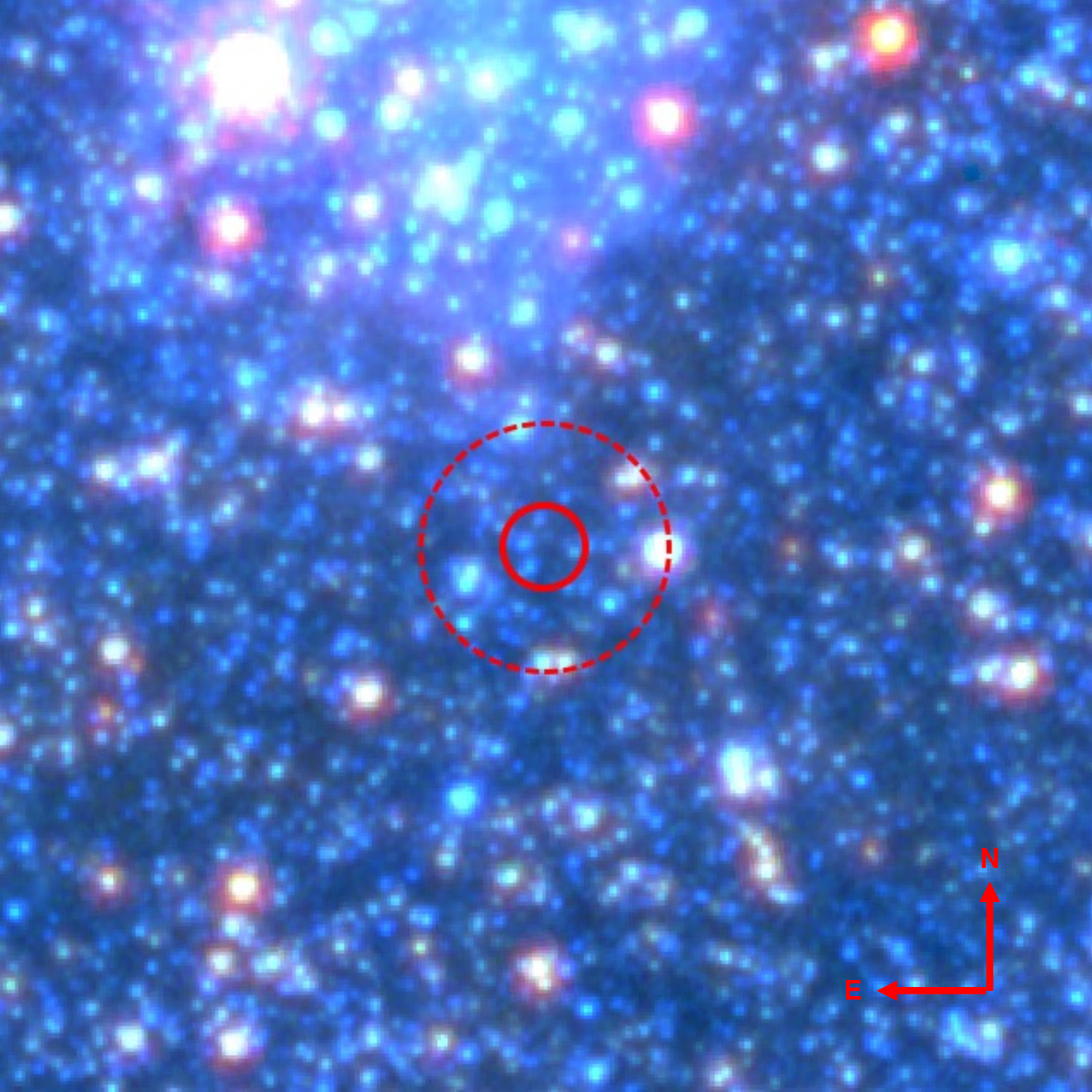}
\caption{Stacked $10'' \times 10''$ ($\sim 40 \times 40$ pc at the distance of M33) HST image with the red F160W, green F814W, and blue F475W  bands used to check for optical counterparts to the new candidate transient 013347.76+303300.0. The $1\sigma$ and $3\sigma$ error circles ($0.4''$ and $1.2''$ respectively) for the X-ray detection are shown, and no bright blue star or background galaxy is seen. The same visual inspection was performed for the other seven new candidate transients as discussed in Section \ref{section:newdetections}.}
\label{fig:hst_stamp}
\end{figure}

We checked the positions of each source in SIMBAD as well as by eye in optical HST imaging \citep{PHATTER2021}, and found no bright blue stars \edit1{(F475W magnitude $< 23$)} or background galaxies in seven of these positions (see Fig. \ref{fig:hst_stamp} for an example). However, 1.16 arcsec away from 013420.89+304947.9 is located a blue supergiant IFM-B 1809 reported in \cite{1993ApJS...89...85I}. This star is within $3\sigma$ of our source position. The proximity and classification of the star make it a strong HMXB candidate if it is the true counterpart to the \edit1{candidate} transient X-ray source we detected. Spectroscopic follow-up of this source is under preparation by another team to classify the companion and confirm its association with the \edit1{candidate} X-ray transient (M. Lazzarini, private communication). 

The lack of optical counterparts for the remaining seven \edit1{candidate} transients may suggest outbursts arising from LMXB systems, where the companion is a low-mass star that would be too faint to detect in current optical surveys. The low count rates of each \edit1{candidate} transient detection are consistent with this idea, as the brightest new candidate transient has luminosity of order $10^{36}$ erg s$^{-1}$. If these detections are true transient sources, they could belong to the faint and very faint X-ray transient classes which are both phenomenological classes of likely LMXB systems \citep{Bahramian2021}. Many previous studies of extragalactic LMXBs focus on sources in globular clusters \citep[e.g.][]{Bildsten2004, Hunt2023}. These sources are not, however, expected to be observed as transient X-ray sources since they are mostly in compact orbits that enable persistent accretion \citep{Padilla2023}. We checked the candidate transient source positions against the PHATTER Star Cluster Catalog \citep{Johnson2022} to see if any sources might be associated with a globular cluster position, and did not find any associations within $5\sigma$ of the X-ray source positions. These sources could therefore be field LMXBs, which is indeed where most transient XRBs are expected to be found if they form according to the "standard" binary formation model (see \cite{Brassington2012} and references therein). A handful of transient X-ray sources in M33 have been reported by previous surveys (see Section \ref{subsec:simdutycycles}), most of which are identified as likely HMXBs \citep{W08}. Validating any of the new candidate transients presented here as LMXBs with deeper X-ray monitoring observations would constitute a significant increase to the fraction of LMXB transients in M33.

\begin{deluxetable*}{ccccccc}

\tablecaption{Summary of measured counts for new candidate transients detected in this survey. The ObsID of the strongest detection and the 0.35---8.0 keV full energy band PNS and net counts are listed for each source. The narrow energy band with the strongest detection is given with the extracted counts. The minimum flux ratio F$_{max}$/F$_{lim}$ between the $1\sigma$ lower limit on the strongest detection count rate and the \citetalias{T11} survey sensitivity is given in the last column. The three sources with F$_{max}$/F$_{lim} > 20$ also exceed a minimum flux ratio of five when calculating this value with the lower limit on the maximum detected count rate.} \label{table:new_transients}

\tablehead{\colhead{Source Name} & \colhead{ObsID} & \colhead{PNS} & \colhead{0.35---8.0 keV Net Cts} & \colhead{Band (keV)} & \colhead{Band Net Cts} &\colhead{F$_{max}$/F$_{lim}$}}
\startdata
013314.01+303839.0 & 23607 & 1.7e-10 & $11_{-3}^{+5}$ & 1.1---2.6 & $7_{-3}^{+4}$ & 112.7 \\
013324.37+303323.0 & 23605 & 2.4e-05 & $5_{-2}^{+3}$ & 1.1---2.6 & $3_{-2}^{+3}$ & 36.6 \\
013347.76+303300.0 & 23605 & 9.8e-06 & $3_{-2}^{+3}$ & 1.1---2.6 & $2_{-1}^{+3}$ & 19.4 \\
013353.05+304710.4 & 23608 & 6.4e-05 & $3_{-2}^{+3}$ & 0.35---1.1 & $3_{-2}^{+3}$ & 18.9 \\
013357.09+304621.7 & 23609 & 4.7e-04 & $2_{-1}^{+3}$ & 0.35---1.1 & $2_{-1}^{+3}$ & 9.9 \\
013406.22+304042.9 & 23609 & 3.7e-08 & $5_{-2}^{+3}$ & 1.1---2.6 & $4_{-2}^{+3}$ & 40.1 \\
013416.32+304319.2 & 23609 & 3.3e-06 & $3_{-2}^{+3}$ & 2.6---8.0 & $2_{-1}^{+3}$ & 19.5 \\
013420.89+304947.9 & 23610 & 3.4e-04 & $3_{-2}^{+3}$ & 0.35---1.1 & $3_{-2}^{+3}$ & 18.0
\enddata
\end{deluxetable*}

\section{Discussion}\label{section:discussion}
We now discuss how we use our catalog of bright X-ray sources to characterize the amplitudes of the changes in detected X-ray emission within our survey and across archival observations, simulate DCs which are consistent with the observed light curves, and discuss \edit1{the variability results and hardness ratios of sources which} have counterparts already identified as candidate and confirmed HMXBs in the literature. 
   
\subsection{X-ray Variability Amplitudes}\label{subsec:variability}

We look for variability on two timescales: short-term (shorter than the roughly seven month baseline of our survey) and long-term ($\sim$1 year or longer). Thus we consider only our survey data when investigating short-term variability, and compare to previous catalogs to investigate long-term variability. 

\begin{figure*}
\centering
\includegraphics[scale=.7]{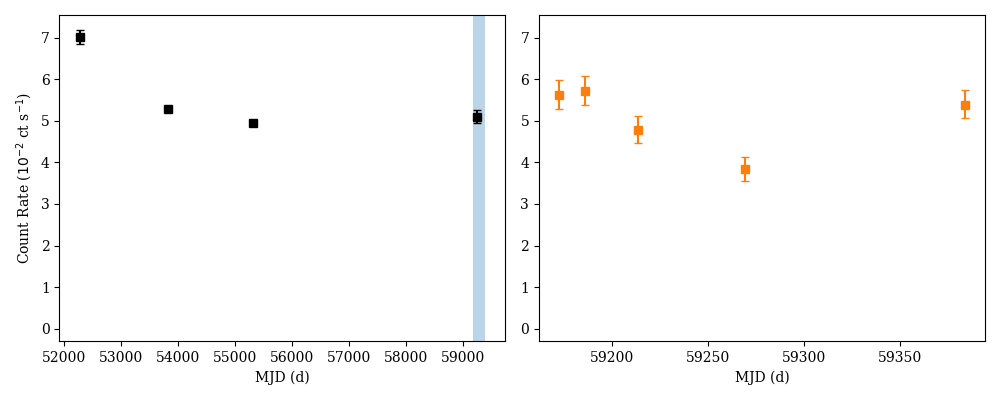}
\caption{X-ray light curve of the HMXB M33 X-7, Source 013334.14+303211.1 in our catalog.}
\label{fig:013334.14+303211.1_lc}
\end{figure*}

\begin{figure}[h]
\centering
\includegraphics[scale = .4]{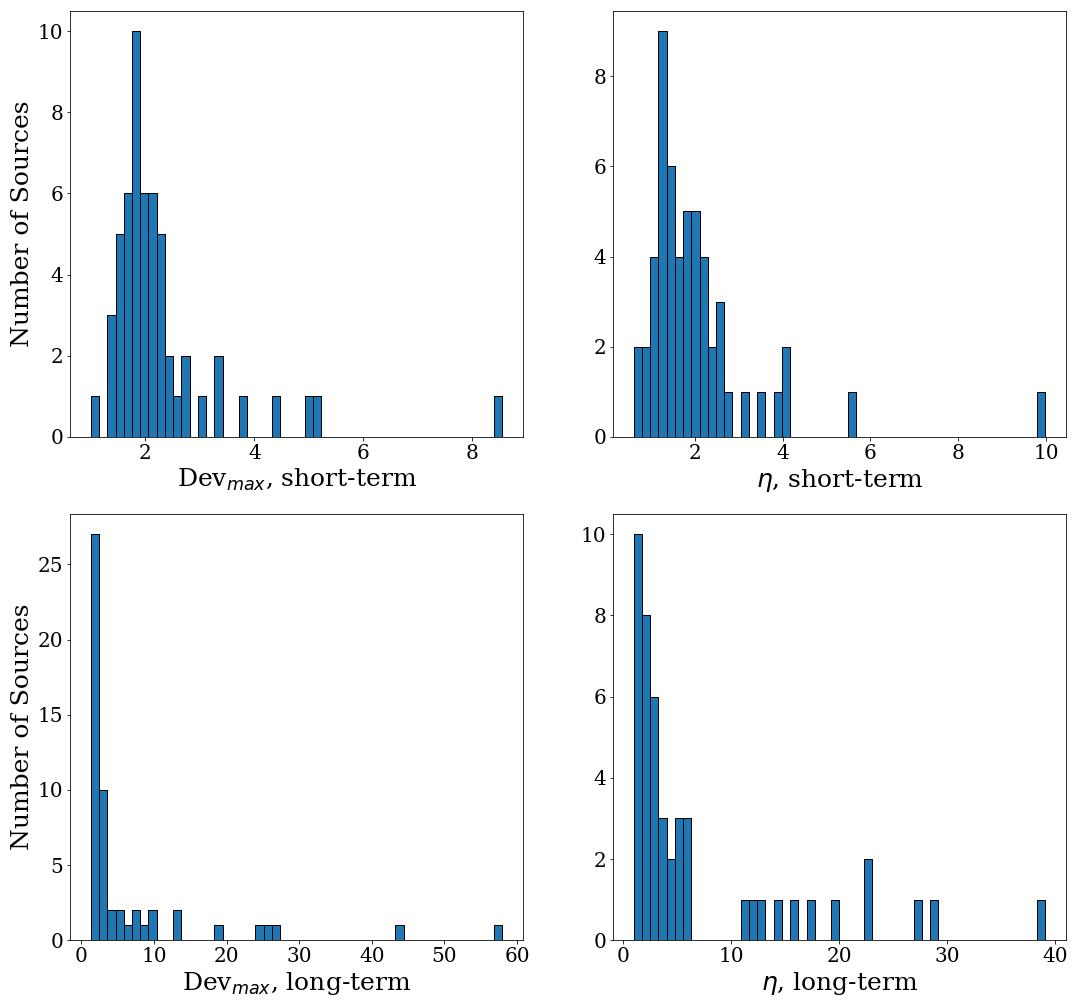}
\caption{Distribution of sources by variability indices Dev$_{max}$ (left) and $\eta$ (right). The top panels show the distributions based on short-term variability (i.e. within our ten ObsIDs covering $\sim$seven months) and the bottom panels show the distributions based on long-term variability over both our observations and archival surveys which cover a maximum baseline of 21 years.}
\label{fig:variability}
\end{figure}

We first consider source variability on short timescales. To quantify variability, we define the variability amplitude Dev$_{max}$ for each source as the maximum difference from the mean count rate divided by the error bar of the measurement most different from the mean count rate in the full 0.35-8.0 keV energy band. We calculate the maximum difference from the mean count rate as $\Delta F_{max} =$ max$(|F - F_{mean}|)$ where $F$ is the 84\% upper or lower limit on each count rate measurement (corresponding to a Gaussian $1\sigma$ limit), depending on whether the measurement is above or below the mean count rate. The mean count rate $F_{mean}$ is weighted by the inverse of the variance. We calculate the variability amplitude as Dev$_{max} = \Delta F_{max} / \sigma$, where $\sigma$ is the upper error bar on the measurement that determines $\Delta F_{max}$ when the maximum measured flux is higher than the mean count rate (such as in Fig. \ref{fig:013313.95+303838.2_lc}), or the lower error bar for sources with a "maximum" measured flux lower than the mean count rate (such as in Fig. \ref{fig:013334.14+303211.1_lc}). We also consider the variability index definition of $\eta = (F_{max} - F_{min}) / \sqrt{(\sigma_{low}^{max})^2 + (\sigma_{upp}^{min})^2}$ used by \citetalias{T11}, where $\sigma$ is the upper or lower error bar on the minimum and maximum measured count rates. The resulting distribution of variability statistics based on our survey observations alone is given in the top panels of Figure \ref{fig:variability}.

We investigated the long-term variability of these sources by considering the archival count rates for each source. The maximum baseline for a source covered by all available data increases from our survey's 7 month duration to about 21 years. We calculated the same \edit1{Dev$_{max}$ and $\eta$ statistics} described above from the merged-ObsID count rates in our survey combined with the modern Chandra equivalent count rates for archival X-ray flux data in \citetalias{M06}, \citetalias{T11} and \citetalias{W15} (converted as described for the long-term light curves above in Section \ref{subsec:previousxraysurveys}). The distribution of variability statistics for the extended baseline is given in the lower panel of Figure \ref{fig:variability}. 

\begin{deluxetable}{ccccc}

\tablecaption{Sources in our catalog with variability indices $\eta > 3$ and Dev$_{max} > 3$ for long-term variability across archival and current observations and short-term variability within our observations.  \label{table:variable_sources}}

\tablehead{\colhead{} & \multicolumn{2}{c}{Long-term} &  \multicolumn{2}{c}{Short-term}
\\ \colhead{Source Name} & \colhead{$\eta$} & \colhead{Dev$_{max}$} & \colhead{$\eta$} & \colhead{Dev$_{max}$}}
\startdata
013311.76+303843.1 & 39.1 & 58.0 & 3.51 & 3.29 \\
013314.01+303839.0 & -- & 3.32 & -- & 3.32 \\
013323.96+303517.1 & 14.5 & 18.9 & -- & -- \\
013324.47+304401.6 & 19.6 & 43.8 & 9.99 & 8.55 \\
013328.70+302724.4 & 23.1 & 27.2 & 5.58 & 4.41 \\
013330.65+303404.1 & 17.5 & -- & -- & -- \\
013331.24+303333.0 & 15.5 & 24.8 & -- & -- \\
013332.24+303955.4 & 4.73 & -- & -- & -- \\
013333.70+303109.8 & 5.77 & 8.33 & -- & -- \\
013334.14+303211.1 & 11.3 & 12.9 & 4.14 & 5.2 \\
013336.42+303743.1 & 3.25 & 3.42 & -- & -- \\
013339.25+304049.7 & 5.46 & -- & -- & -- \\
013342.02+303848.6 & 22.6 & 5.36 & -- & -- \\
013342.55+304253.5 & 12.3 & -- & -- & -- \\
013343.28+304630.9 & 3.16 & 9.96 & -- & -- \\
013346.55+303748.7 & 5.08 & -- & 3.8 & -- \\
013348.94+302946.9 & -- & 3.36 &--  &  --\\
013350.49+303821.2 & 12.9 & 10.1 & -- & -- \\
013352.26+302707.4 & -- & 4.63 & -- & -- \\
013356.76+303729.4 & -- & 3.0 & 4.1 & 3.85 \\
013356.82+303707.4 & 3.46 & 3.61 & -- & -- \\
013357.21+305133.8 & 4.08 & -- & -- & -- \\
013358.80+305004.3 & 3.52 & -- & -- & -- \\
013401.14+303136.6 & 6.32 & 6.18 & -- & -- \\
013402.02+303004.7 & 5.07 & 13.7 & -- & -- \\
013410.52+303946.2 & 3.11 & 7.72 & -- & -- \\
013425.68+305518.1 & 27.6 & 25.5 & -- & -- \\
013426.43+304446.4 & -- & 5.63 & -- & 5.05 \\
013426.93+304313.2 & -- & 7.21 &--  & -- \\
013439.93+305143.5 & 6.16 & -- & -- & -- \\
013445.07+304924.5 & 3.38 & 3.11 & -- & -- \\
013451.72+304618.0 & 29.0 & -- & 3.12 & 3.01
\enddata

\end{deluxetable}

The top panels of Figure \ref{fig:variability} shows that most sources in our catalog exhibit only low amplitude variability within the seven month duration of our observations, with peaks at \edit1{Dev$_{max} = 2$ and $\eta = 2$}. The same is true for the extended baseline variability statistics shown in the lower panels of Figure \ref{fig:variability}, although several sources vary with much higher significance than in the short-term calculations. \edit1{The distribution of Dev$_{max}$ for both short-term and long-term data falls off} \edit1{at Dev$_{max} = 3$}, thus we chose to define sources with \edit1{Dev$_{max} > 3$} as significantly variable. Within our observations only, eight of our sources pass this threshold, while 23 of our sources pass the threshold when including archival data, showing that significant variability is more common at longer timescales than shorter ones in our sample. The sources displaying variability amplitudes greater than \edit1{Dev$_{max} = 3$} on long- or short-timescales are given in Table \ref{table:variable_sources}.

The distribution of $\eta$ for short-term variability also peaks just below $\eta = 2$ and has a tail that drops off considerably by $\eta = 3$, which we again use as the threshold for flagging sources as significantly variable according to this test. Seven sources pass this cut for the short-term calculation of $\eta$. For the long-term calculation, the $\eta$ distribution has a much longer tail. Nearly half of the sources in our catalog (26 sources) pass the threshold for variability $\eta > 3$ on the extended baseline, 18 of which also satisfy $\eta > 5$ which is the threshold that \citetalias{T11} uses to flag sources likely to be significantly variable. A total of 27 sources are indicated as variable on at least one timescale according to the thresholds for $\eta$.

Our definition of the variability index Dev$_{max}$ is sensitive to a different variability pattern than the index $\eta$ used by \citetalias{T11}, but still identifies similar numbers of variable sources among the bright X-ray source population in the direction of M33. For long-term variability, 32 sources are indicated to be variable by at least one of the two statistics, with 17 of these sources flagged by both Dev$_{max}$ and $\eta$. For the short-term variability within our survey's duration, nine sources are flagged as highly likely to be variable with at least one index, with agreement from both on six sources.

One notable source is 013311.76+303843.1. In the top left panel of Figure \ref{fig:variability}, this source appears at the very start of the tail with a value of \edit1{Dev$_{max} = 3.29$}. The variability increases significantly to \edit1{Dev$_{max} = 58$} when considering literature values. The error bars for both literature measurements and in our observations have little to no overlap with each other except in four of our observations for which the observed count rate is essentially constant. Thus, the calculated variability amplitudes provide a robust description of the true variability of the source. For 013311.76+303843.1 it is likely that the fluctuations in count rate in our observations are part of a more lengthy variability cycle. 

Sources 013343.28+304630.9 and 013342.02+303848.6 are known variable foreground stars in the direction of M33 \citep{Hatzidimitriou2006, T11}, which were also the only sources identified as foreground stars in \citetalias{T11} detected in our survey. \edit1{Our variability statistics both find that these two sources are significantly variable on the extended baseline (see Table \ref{table:variable_sources}).} Without a known nonvariable foreground source to compare to, it is difficult to define a threshold for variability intrinsic to sources in M33. Even so, a more conservative threshold of \edit1{$\eta > 5$} yields 18 sources with significant variability on at least one timescale.

The numerous nonvariable sources in our catalog serve as an additional validation tool for the data reduction techniques used for this survey. If the images from each ObsID are well aligned, then sources that are persistent across timescales should show low variability within our observations and when including the archival observations. There are \edit1{15} sources that show persistent flux within our observations and long-term, and \edit1{23 sources that do not vary over the duration of our observations but do vary long-term according to at least one variability index threshold}. Since \edit1{over one quarter} of the sources in our catalog are persistent, it is likely that our images are well aligned and the 23 sources varying only long-term are sources with variability cycles extending beyond the duration covered by our survey.

The fraction of persistent sources in our catalog is consistent with trends observed in the Milky Way and Magellanic Clouds. Two thirds of Galactic X-ray binaries are persistent sources \citep{1996ARA&A..34..607T}. The recurrence timescale of X-ray transients is on the order of years \citep{Degenaar12}, which is longer than the duration of this survey but encompassed by the sources showing long-term variability with the extended baseline. Supergiant fast X-ray transients, a subclass of HMXBs, flare on timescales of a few ks repeatedly over timescales of days to months \citep{Sidoli2019}. Although these sources reach relatively low peak X-ray luminosities (rarely exceeding $L_X \sim 10^{36}$ erg s$^{-1}$), the strongest of these types of sources could be detectable in our survey. 

\subsection{Duty Cycle Constraints}\label{subsec:simdutycycles}

A source’s DC describes the fraction of time it is in a high emission state. \cite{Sidoli2018} found that each HMXB subclass typically follows a different range of DCs, with soft X-ray transients displaying low DCs ($< 5\%$) and wind-fed HMXBs with supergiant companions displaying DCs above 10\%. To constrain the X-ray DCs of sources in our catalog, we simulated light curves for a grid of DCs and timescales and determined the frequency with which the observed light curves match the simulated light curves. This duty cycle analysis can only be performed for sources that are considered "off" (by our criteria described below) in at least one observation, which is a sub-sample of the total population detected and analyzed in this work.

We attempted to constrain the DCs of sources in our catalog which were not detected in all observations that covered their positions. If the X-ray sources in the direction of M33 all had 100\% DCs, we would expect to detect all 94 \citetalias{T11} sources covered by our survey region that were previously above our $10^{-4}$ ct s$^{-1}$ detection limit and did not have very soft $\Gamma > 3$ spectral index fits (which our survey would likely not detect due to the deterioration of the ACIS instrument’s soft-band sensitivity; see Section \ref{subsec:previousxraysurveys}). Instead, we detected 48 of these sources. Thus, about half of these sources, many of which are expected to be HMXBs, are likely variable on long timescales. The archival survey luminosity limits in Table \ref{table:upperlims} show that previous X-ray catalogs would have detected any sources in our catalog if they had been in outburst during those epochs. 

To simulate the short-term variability within the duration of our survey ($\sim$ seven months), we produced 100 simulated light curves for each combination of DCs and periods of variability (which we refer to as "timescale"). The DC grid ranged from 5-100\% at 5\% intervals and the timescale grid included 10-200 day timescales at 10 day intervals. Each simulated light curve is a series of quasi-periodic cycles in which the source is assigned as “on” (in a high-emission state) for the number of days equal to the cycle length multiplied by the DC. All days the source is on in each cycle are sequential, and the day in the cycle that the source turns on is random. The time between our earliest observation in the survey presented here to the end of our observations is 228 days. We determined the number of cycles needed to construct each curve by dividing the total observation time by the timescale and rounding up, then adding one extra cycle, allowing us to randomize the cycle start time.

\begin{figure}[h]
\centering
\includegraphics[scale=.55]{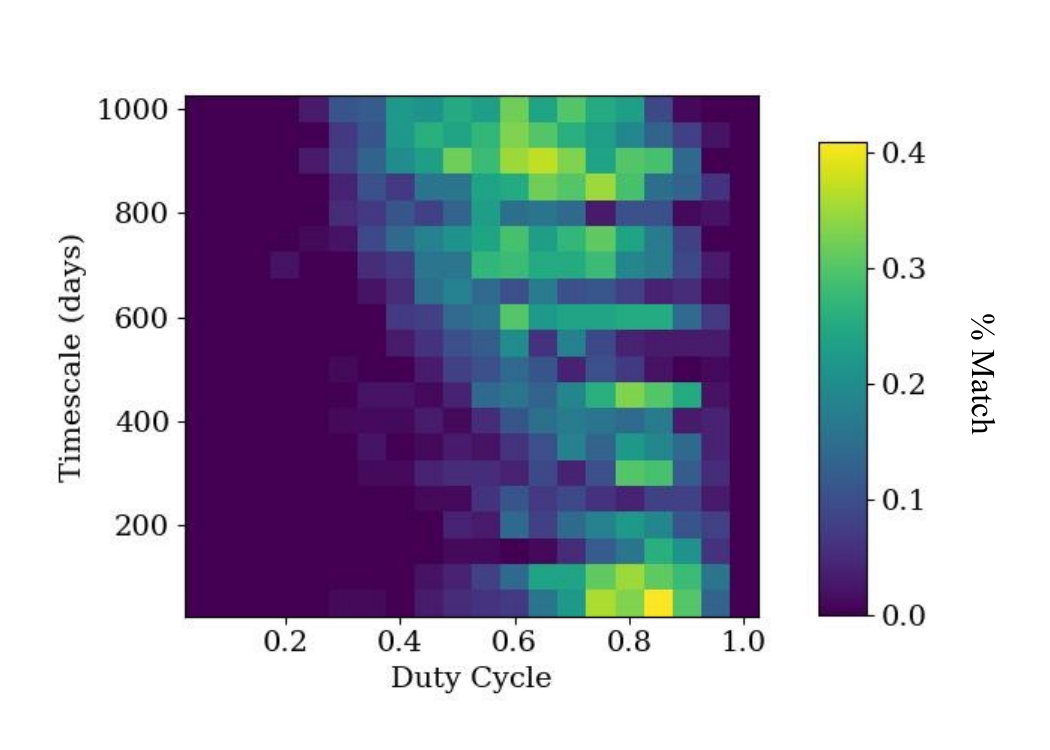}
\caption{Representative duty cycle simulation result for a source (013314.01+303839.0) marked as on in all our observations and off in at least one of \citetalias{M06}, \citetalias{T11}, and \citetalias{W15}. The plotted quantity is the fraction of simulated light curves that match the observed light curve for each combination of DC and variability timescale. The DCs for these sources are constrained to be greater than 30\% with good agreement across all timescales we could probe within our extended baseline.}
\label{fig:dutycycle_longex}
\end{figure}

We then mapped our observed light curves to a modified light curve to directly compare to the simulations, marking the source as “on” or “off” (in a low-emission state) for each ObsID. We define a source to be off if the upper error is lower than 1\% of the lower error on the faintest detection. The 1\% buffer region is included to mark a single observation of source 013350.49+303821.2 as on so that its DC may be constrained, despite the upper error being slightly below the faintest detection. Next, we compared the observed light curves to the simulated curves by randomly choosing a day in the first cycle of the simulated curve to set as the first day. We considered a simulated light curve to match an observed curve if the source activity was consistent on all days of our observations. For each combination of timescale and DC, we recorded the fraction of simulated curves that matched each source. The results are represented as 2D histograms as in Figure \ref{fig:dutycycle_longex}.

\begin{figure}[h]
\centering
\includegraphics[scale=0.4]{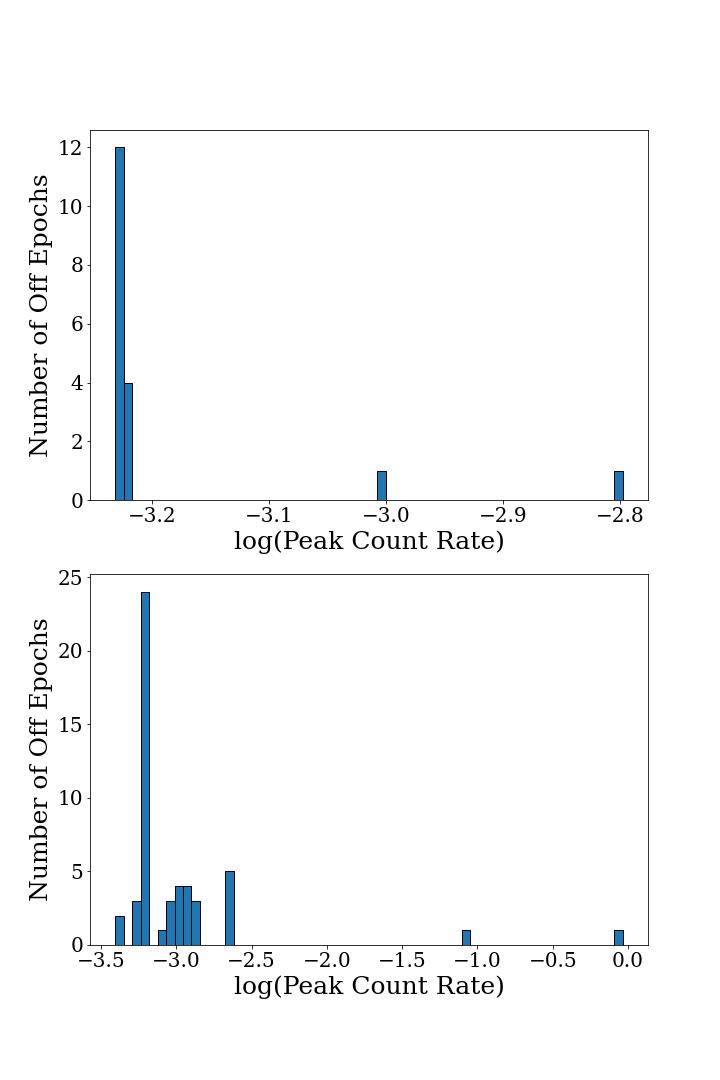}
\caption{Number of epochs sources are labelled as off binned by the log of the peak count rate in ct s$^{-1}$. The top panel shows the number of off epochs for short-term light curves. The lower panel shows the number of off epochs for long-term light curves including our observations and archival data.}
\label{fig:peakcountrate}
\end{figure}

The distribution of the number of off epochs by peak count rate is shown in Figure \ref{fig:peakcountrate}. Within our survey, only 11\% of our sources have any off epochs. When including literature results, still only 34\% of our sources have any off epochs. To further break down the variability of the observed sources, we compare the number of off epochs observed to the peak count rate recorded. In our observations, 28\% of sources with peak count rates less than or equal to $10^{-3}$ ct s$^{-1}$ (corresponding to L$_X \lesssim 1.7 \times 10^{36}$ erg s$^{-1}$ using the same assumptions to derive our limiting luminosity in Section \ref{section:results}) have at least one off epoch and 22\% have at least four off epochs. For peak count rates greater than $10^{-3}$ ct s$^{-1}$ and less than or equal to $10^{-2.5}$ ct s$^{-1}$ ($1.7 \times 10^{36}$ erg s$^{-1} <$ L$_X < 5.3 \times 10^{36}$ erg s$^{-1}$), 4\% of sources have at least one off epoch and none have more than three. There are also no sources with peak count rates at values greater than $10^{-2.5}$ ct s$^{-1}$ and any off epochs. The decreasing number of sources with off epochs as peak count rate increases \edit1{suggests} that faint sources are more likely to turn off than bright sources, \edit1{although fluctuations near our survey sensitivity limit may exaggerate this trend.} When including past survey measurements, 93\% of sources with a peak count rate less than or equal to $10^{-3}$ ct s$^{-1}$ have at least one off epoch, and 29\% have at least four. For peak count rates greater than $10^{-3}$ ct s$^{-1}$ and less than or equal to $10^{-2.5}$ ct s$^{-1}$, 21\% of sources have at least one off epoch and none have greater than three. There are no sources with any off epochs at peak count rate values larger than $10^{-2.5}$ ct s$^{-1}$. The long-term data also supports the trend of the number of off epochs decreasing as peak count rate increases. The larger percentage of sources with off epochs including long-term observations \edit1{is also consistent with the finding from the variability statistics in Section \ref{subsec:variability}} that sources in our data set tend to vary more long-term than short-term. 

We next simulated additional light curves with an extended baseline to match to the long-term X-ray light curves including our observations and those from \citetalias{M06}, \citetalias{T11} and \citetalias{W15}. The time from the earliest survey observation to our last observation is just over 7646 days, so we used 50-1000 day timescales with 50 day steps, and 5-100\% DCs in increments of 5\%. to create the modified on-off light curves of our observations and archival observations, we considered the source on for all individual observation days in the archival surveys where the source was detected. We checked for matches to transient sources in the archival surveys for which the assumption of being on at all survey times would not be valid. Only one source in the match to \citetalias{M06} (their Source Number 251, our source 013431.97+303454.2) was flagged as having both dropped below their detection threshold and being significantly variable. Our catalog includes one additional match to a transient source identified from the \citetalias{T11} data set by \cite{W08}, designated as XRT-2 by that paper (\citetalias{T11} Source Number 210, our source 013332.24+303955.4). \citetalias{W15} detects XRT-2 but does not flag it as transient since it was a faint source that did not have to vary by a factor $> 10$ to explain the past nondetections. We discuss the potential impact on our DC results for these two known transient sources below.

We matched our observed long-term light curves to the new simulated curves, considering a match to be where source activity was consistent for all days in our observations and at least one day in each literature survey. Most sources had the highest fraction of simulated light curves in agreement with the observed light curves for shorter timescales, from 50 to 200 days. Thus, we created a new set of simulated curves with timescales every 10 days from 10 to 200 days and compared to the data again. 

We performed this DC analysis for the 19 sources in our catalog which could be classified as off in at least one observation when including archival data. Of these, strong constraints can be made on six sources that were definitely not detected in at least one observation within our survey. The other 13 sources were on in all of our relevant observations and only off in at least one of \citetalias{M06}, \citetalias{T11}, \citetalias{W15} (six of which are new \edit1{candidate} transients presented in Section \ref{section:newdetections}). These 13 sources are consistent with DCs $> 30\%$ without any strong constraint on timescale (see Fig. \ref{fig:dutycycle_longex}), so we expect these sources to have DCs that operate on timescales longer than the seven month duration of our survey. This is consistent with common intervals between XRB transient outbursts of multiple years \citep{Degenaar12}. These 13 sources with DCs $>30\%$ that operate on timescales longer than those probed in this monitoring survey are 013314.01\allowbreak+303839.0, 013324.37\allowbreak+303323.0, 013336.04\allowbreak+303333.1, 013337.94\allowbreak+303837.4, 013356.09\allowbreak+303024.8, 013357.09\allowbreak+304621.7, 013358.03\allowbreak+303201.0, 013358.09\allowbreak+303438.0, 013401.12\allowbreak+303242.2, 013402.87\allowbreak+304151.4, 013406.22\allowbreak+304042.9, 013416.32\allowbreak+304319.2, and 013420.89\allowbreak+304947.9.

\begin{figure*}
\includegraphics[width=\textwidth]{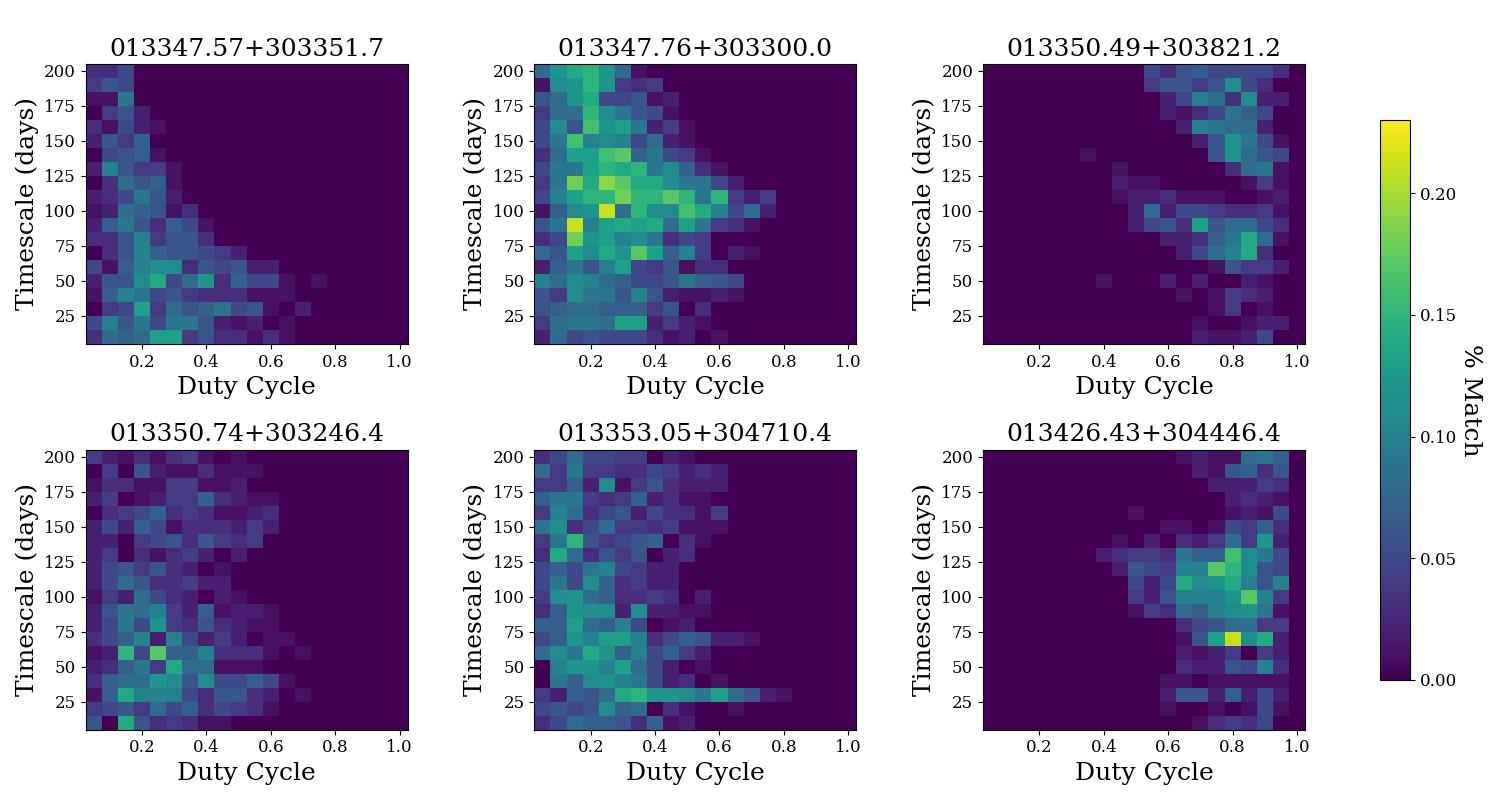}
\caption{Simulated duty cycle match grids for the sources with at least one "off" epoch within our survey.}
\centering
\label{fig:dutycycle_hist}
\end{figure*}

For the six sources which have at least one off observation in our survey and an archival survey, the fraction of simulations that match the observed light curves are shown in the density plots in Figure \ref{fig:dutycycle_hist}. We report the timescale and DC most consistent with the observed light curves as well as the strength of the peak for these sources in Table \ref{table:dc_summary}. The upper and lower errors are defined by the highest and lowest DC and timescale that match the observed light curves in at least 1\% of the simulations. Of these six sources, two have DCs above 50\% (013350.49+303821.2 and 013426.47+304446.4). Source 013426.43+304446.4 varies significantly over our observations, while both sources with DCs higher than 50\% meet the criteria for being significantly variable when considering the survey observations each was present in. Source 013350.49+303821.2 was on the detector for all ten of our observations and 013426.43+304446.4 for eight. Meanwhile, the remaining four sources were only on the detector for five observations each. Two of these sources are new \edit1{candidate}  transients, 013347.76+303300.0 (see Fig. \ref{fig:013347.77+303259.8_lc}) and 013353.05+304710.4, and are measured as being on in only one observation, indicating very short outburst lengths. Although these sources have lower DCs, these small outbursts indicate they may be transients with shorter cycle durations. 

Our survey detects the known transient \citetalias{M06} Source Number 251 as on in all four observations where it was on the detector. This source is marked as on in each of the archival surveys as well, so no DC information could be obtained when using the assumption that the source was on during all days in the archival surveys. We ran an additional set of duty cycle simulations for this source which included the \citetalias{M06} nondetection with an upper limit below the observed flux of the source as an off epoch and found that the DC must be higher than 30\% with agreement across timescales, similar to the 13 sources discussed above. We also detect the known transient XRT-2, which is listed as an HMXB candidate by \citetalias{T11} and \citetalias{G18} (\citetalias{T11} Source 210; see Section \ref{subsec:hmxbs} below). We ran an additional suite of duty cycle simulations for this source including the off epoch within the \citetalias{T11} ObsIDs and obtained the same result of DC $> 30\%$.

\begin{deluxetable}{cccc}

\tablecaption{Results from the most successful duty cycle simulation constraints. Source 013350.49+303821.2 had two equally likely simulated eruption timescales (70 d and 80 d) with the same duty cycle, indicating that its best match was on the edge of the pixel at 75 d. The upper and lower errors are defined by the highest and lowest timescale/duty cycle values where more than 1\% of the simulated light curves matched the observed light curve. \label{table:dc_summary}}

\tablehead{\colhead{Source Name} & \colhead{Peak Match (\%)} & \colhead{Timescale (d)} & \colhead{DC (\%)}}
\startdata
013347.57+303351.7 & 14 & $50_{-40}^{+10}$ & $25_{-5}^{+15}$ \\
013347.76+303300.0 & 23 & $110_{-100}^{+90}$ & $50_{-40}^{+10}$ \\
013350.49+303821.2 & 14 & $75_{-5}^{+115}$ & $85_{-15}^{+0}$ \\
013350.74+303246.4 & 17 & $60_{-50}^{+20}$ & $25_{-10}^{+15}$ \\
013353.05+304710.4 & 17 & $110_{-100}^{+30}$ & $20_{-10}^{+40}$ \\
013426.43+304446.4 & 21 & $70_{-0}^{+70}$ & $80_{-15}^{+10}$
\enddata

\end{deluxetable}

\edit1{We identify 17 sources with DC $>30\%$ out of the 21 sources with at least one off epoch.} Observed HMXB populations in both the Milky Way and Magellanic clouds \edit1{generally show low fractions of high DC sources}. Of the sample of 56 sources observed in the Milky Way by \textit{INTEGRAL}, \cite{Sidoli2018} reports only 23\% as having DCs above 25\%. Similarly, \cite{WATCHDOG} report the majority of observed transient black hole XRBs in both the Milky Way and Magellanic Clouds to have DCs less than 10\% (most of these are LMXBs). About 75\% of BH-LMXBs in the Milky Way turn on once every 50 years \citep{Mori2021}. \edit1{We can identify high DC sources with the present data set, but more frequent X-ray monitoring would likely enable constraints on many more low-duty DCs.}

\subsection{Confirmed and Candidate HMXBs}\label{subsec:hmxbs}

Our catalog includes observations of 26 confirmed and candidate HMXBs previously identified in the literature by \citetalias{T11}, \citetalias{G18}, \cite{Yang2022} and \citetalias{Lazzarini2023}. These are listed in Table \ref{table:hmxb_names} and \edit1{included in the full version of Table \ref{table:final_catalog_summary} available online,} including the nucleus source M33 X-8 (013350.88+303936.5) and the eclipsing HMXB M33 X-7 (013334.14+303211.1; see Fig. \ref{fig:013334.14+303211.1_lc}). Every HMXB candidate is associated with a \citetalias{T11} source by the study that identifies them. We find that 17 of these known HMXB candidates show significant variability on at least one timescale. Six of these sources are persistent within our survey and only off in an archival survey, consistent with the DCs constrained to $>30\%$ discussed above in Section \ref{subsec:simdutycycles}. The remaining eight sources show persistent emission.

\begin{deluxetable}{ccccc}
\tablecaption{HMXB candidates previously identified in the literature with matches to sources in our catalog. \label{table:hmxb_names}}

\tablehead{\colhead{Source Name} & \colhead{T11\tablenotemark{a}} & \colhead{G18\tablenotemark{b}} & \colhead{Y22\tablenotemark{c}} & \colhead{L23\tablenotemark{b}}}
\startdata
013324.47+304401.6 & 158 &  & 3 &  \\
013328.70+302724.4 & 180 &  & 2 &  \\
013330.65+303404.1 &  &  & 10 &  \\
013332.24+303955.4 & 210 & 210 &  &  \\
013333.70+303109.8 &  &  & 9 & 221 \\
013334.14+303211.1 & 225 & 225 & 4 &  \\
013336.04+303333.1 &  & 237 &  & 237 \\
013337.94+303837.4 &  & 250 &  & 250 \\
013340.04+304323.0 &  & 268 &  & 268 \\
013342.55+304253.5 & 281 & 281 & 18 & 281 \\
013346.55+303748.7 &  &  & 15 & 299 \\
013350.49+303821.2 &  & 316 &  & 316 \\
013350.88+303936.5 & 318 & 318 & 1 & 318 \\
013356.76+303729.4 &  & 347 & 20 & 347 \\
013356.82+303707.4 & 348 & 348 &  & 348 \\
013358.09+303438.0 &  & 358 &  & 358 \\
013358.80+305004.3 &  &  & 23 &  \\
013401.12+303242.2 &  & 384 &  & 384 \\
013402.87+304151.4 &  & 398 &  & 398 \\
013407.80+303553.8 &  &  &  & 415 \\
013410.52+303946.2 &  & 424 &  & 424 \\
013425.68+305518.1 &  &  & 6 &  \\
013426.43+304446.4 &  &  &  & 497 \\
013426.93+304313.2 &  &  &  & 502 \\
013436.36+304713.9 &  &  & 13 &  \\
013445.07+304924.5 & 589 &  &  &
\enddata

\tablenotetext{a}{T11 source number for the sources identified in that catalog as XRBs or XRB candidates only. Note that all 26 HMXB candidates have matches to T11 sources even if they are not identified by T11 as candidate XRBs.}
\tablenotetext{b}{G18 and L23 sources are identified by their T11 source numbers.}
\tablenotetext{c}{Y22 ID number.}

\end{deluxetable}

\citetalias{G18} identified 55 HMXB candidates from a combination of the \citetalias{T11} catalog and archival HST and Spitzer imaging. Fourteen of the \citetalias{G18} candidates were observed in our survey. There is one source in the \citetalias{G18} list that we do not detect at all in our survey (J013410.69+304224.0) that falls within our survey region and was above our detection limit in the past, which was classified as a supernova remnant by \cite{Long2010}. 

Twelve of the 28 sources from the \cite{Yang2022} candidate list observed with NuSTAR are detected in our survey. The positions of these sources on a hardness-intensity diagram in that paper (Figure 3) indicate the nature of the compact object in each XRB candidate. Most of our detected sources with matches to the \cite{Yang2022} catalog are identified as likely containing a BH, with the exceptions of \edit1{013328.70+302724.4 (M33 X-6; Y22 Source 2),} 013342.55+304253.5 (Y22 Source 18) and possibly 013436.36+304713.9 (Y22 Source 13) which are more consistent with where the pulsar population lies.

Our survey detects 17 of the 65 HMXB candidates reported in \citetalias{Lazzarini2023}, three of which are classified as HMXB candidates for the first time. Multiple optical counterparts from PHATTER \citep{PHATTER2021} are listed for nine of these sources. Among the top ranked optical counterparts identified by \citetalias{Lazzarini2023} for the 17 sources we detect in this survey, about half are main sequence B stars. This is a lower fraction than in the full \citetalias{Lazzarini2023} catalog, which reports 75\% of top-ranked optical counterparts to \citetalias{T11} sources to be main sequence B stars. We also detect two of the four sources with main sequence O star companions, all three of the sources with giant O star companions, and the single source with a giant B star companion. This is consistent with our systematic limitation of detecting only the most energetic systems with our shallow survey, and we also expect some intrinsic variability due to a few sources being at different phases in their activity cycles during the \citetalias{T11} observations and during our survey. Seven sources of this list are “off” in at least one observation in our survey which suggests that these sources are significantly variable and therefore stronger HMXB candidates. All sources from that list are covered by our survey’s detection area and four of the undetected sources were above our detection threshold in the past (\citetalias{T11} IDs: 274, 277, 391 and 452). T11 ID 233 was also above our detection limit in the past, but was found in \citetalias{T11} to have a soft power law spectrum with $\Gamma = 2.09$ so this source would be below our detection limit due to the deterioration of Chandra's soft-band sensitivity. The other four sources would have been detectable by our survey at their luminosities observed by \citetalias{T11} which suggests that they were in a low X-ray emission state at the times of all of our observations and are also significantly variable sources. T11 277 has a supergiant B star optical companion, and T11 391 has a main sequence B star. T11 274 has two possible optical counterparts which are a main sequence B star and a main sequence O star. T11 452 has three possible counterparts, two of which are main sequence B stars.

Eight of the sources in our catalog with cross matches to \citetalias{T11} are identified there as XRB or related classes. Only one of these does not make it into the three other lists discussed, and was identified as a quasar by \cite{Neugent2011}. \edit1{We include this source (013445.07+304924.5) in Table \ref{table:hmxb_names} for completeness.} We detect all sources marked as XRB or related classes that were above our detection limit in the \citetalias{T11} catalog and covered by our survey region.

\begin{figure*}
\centering
\includegraphics[scale=0.7]{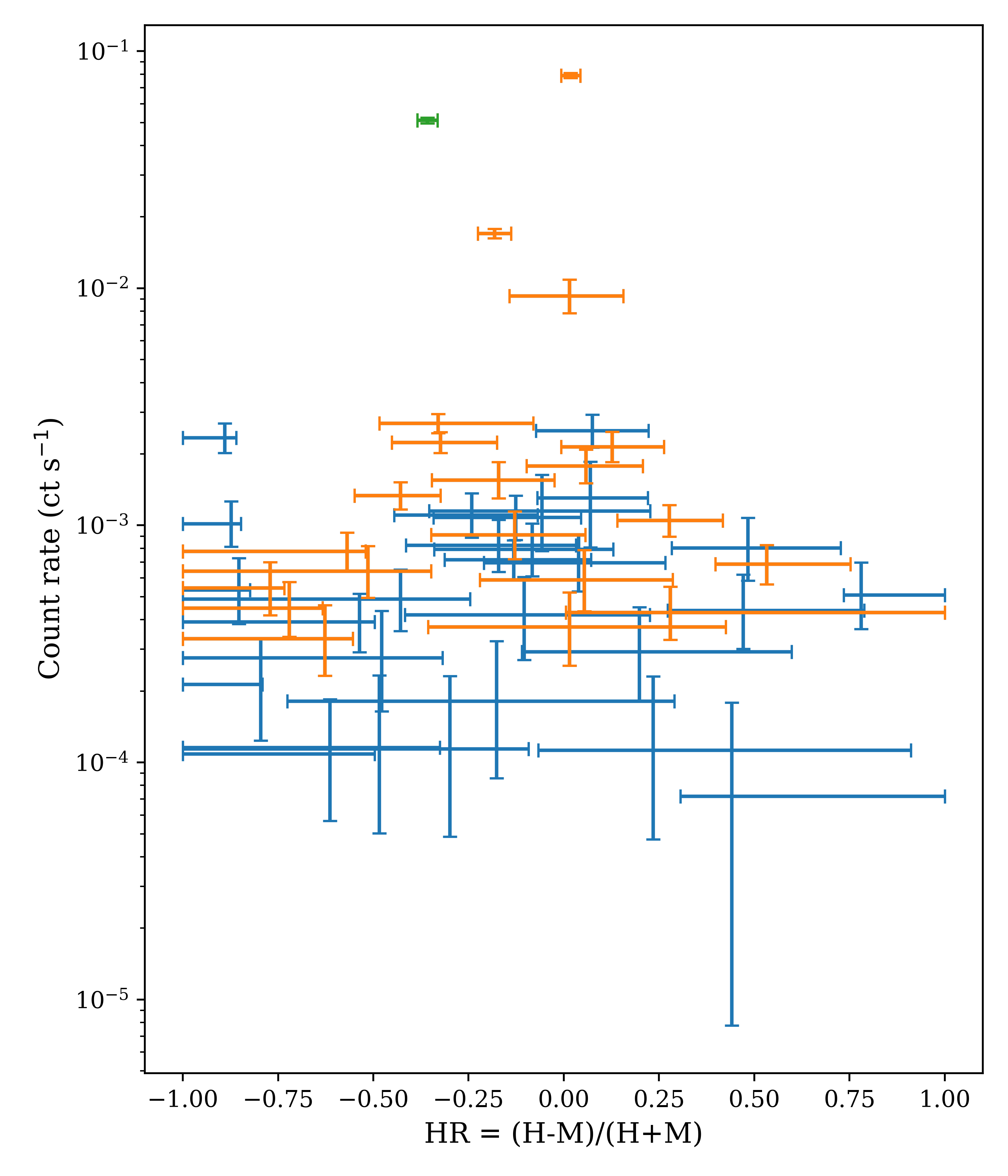}
\caption{Hardness ratio (HR) between the hard 2.6---8.0 keV energy band the medium 1.1---2.6 keV energy band versus merged-ObsID 0.35---8.0 keV count rate. HMXB candidates are plotted in orange, and the well-studied HMXB M33 X-7 is plotted in green. Other catalog sources which we could constrain the HRs for are plotted in blue. Most HMXB candidates are as hard as or harder than X-7.}
\label{fig:hardness_intensity}
\end{figure*}

HMXBs with active accretion are hard X-ray sources \citep{diSalvo2004,Yukita2016,Yang2022}. We attempted to characterize the hardness of all sources in our catalog using the Bayesian Estimation of Hardness Ratios program (BEHR; \cite{Park2006}). Due to the poor soft-band response of the ACIS instrument, hardness ratios (HRs) using counts in the 0.35---1.1 keV energy band are severely biased. We therefore only consider a HR between the medium 1.1---2.6 keV energy band (M) and the hard 2.6---8.0 keV energy band (H). Seven sources resulted in unconstrained HRs due to low counts. We show a plot of HR versus count rate for the 48 sources in our catalog that BEHR found HRs with constraining $1\sigma$ error bars for in Figure \ref{fig:hardness_intensity}. We find that the HRs of 17 out of the 21 HMXBs with constrained HRs are as hard or harder than the well-studied HMXB X-7, indicated in green in Figure \ref{fig:hardness_intensity}. The remaining four HMXB candidates (Sources 013337.94+303837.4, 013340.04+304323.0, 013350.49+303821.2 and 013407.80+303553.8) have HR upper limits that are softer than X-7, but still may not necessarily be soft sources since we cannot calculate a HR with the soft band counts.

\section{Conclusions}

We have presented a new five-epoch survey of 55 bright X-ray sources in M33 probing multiple timescales, including eight previously undetected \edit1{candidate} transients and \knowncandidates{} known HMXB candidates outside of the nucleus. We characterized the short- and long-term variability amplitudes of 32 significantly variable sources, identified 15 nonvariable sources, and found constraints on the DCs of 21 sources in our catalog. We found a higher fraction of sources with DCs $>30\%$ than previous HMXB literature suggests due to the infrequent observation history of M33. There are many observations that would be useful to classify the variable X-ray sources in this work, including further X-ray monitoring and inspection of the optical counterparts to these sources. While a majority of the bright X-ray sources in M33 are predicted to be HMXBs, there are likely also background AGN present in the sample as contaminants. A careful multi-wavelength selection is needed to get the cleanest intrinsic XRB sample for targeted timing studies. Once the population of HMXBs in M33 is well defined, deeper X-ray observations across the entire population are necessary to improve our understanding of these energetic systems as a whole.

\acknowledgements
Support for this work was provided by the National Aeronautics and Space Administration through Chandra Award Number GO1-22080X issued by the Chandra X-ray Observatory Center, which is operated by the Smithsonian Astrophysical Observatory for and on behalf of the National Aeronautics Space Administration under contract NAS8-03060. The scientific results reported in this article are based on observations made by the Chandra X-ray Observatory. This work has made use of SAOImage DS9, developed by the Smithsonian Astrophysical Observatory \citep{DS9}.

This paper employs a list of Chandra datasets, obtained by the Chandra X-ray Observatory, contained in~\dataset[DOI:10.25574/cdc.158]{https://doi.org/10.25574/cdc.158}.

\software{Astropy \citep{astropy:2013, astropy:2018,astropy2022}, Chandra Interactive Analysis of Observations (CIAO; \cite{CIAO}), ACIS-Extract \citep{AE2012}, Bayesian Estimation of Hardness Ratios (BEHR; \cite{Park2006})}

\bibliographystyle{aasjournal}
\bibliography{main}

\end{document}